\documentstyle[12pt,aasms4]{article}
\makeatletter
\def\jnl@aj{AJ}
\ifx\revtex@jnl\jnl@aj\fi
\makeatother

\def \etal {et al.\thinspace}

\def \hii   {\ion{H}{2}}

\def \ne{$N_e$\thinspace }
\def \nd{$N$\thinspace }
\def \te{$T_e$\thinspace }
\def \cm3{cm$^{-3}$\thinspace }
\def \mic   {$\mu$m\thinspace }

\begin{document}

\title{Ionization Structure and Spectra of Iron in Gaseous Nebulae}
\author{Manuel A. Bautista \& Anil K. Pradhan}  
\affil{Department of Astronomy, The Ohio State University}
\affil{174 West 18th Avenue, Columbus, OH\ 43210-1106}

\begin{abstract}

The emission spectra and the ionization structure of the low ionization
stages of iron, Fe~I--IV, in gaseous nebulae are studied.
This work includes: (i) new atomic data: photoionization cross sections,
total e-ion recombination rates, excitation collision strengths,
and transition probabilities calculated under the Iron Project by the
Ohio State atomic-astrophysics group; (ii)
detailed study of excitation mechanisms for the [Fe~II],
[Fe~III], and [Fe~IV] emission, 
and spectroscopic analysis of the observed IR, optical, and UV spectra;
(iii) study of the
physical structure and kinematics of the nebulae and their ionization fronts. 
Spectral analysis of the well observed Orion nebula is carried
out as a test case, using extensive collisional-radiative and 
photoionization models. 
It is shown that the [Fe~II] emission from the Orion nebula is predominantly 
excited via electron collisions in high density partially 
ionized zones; radiative fluorescence is relatively less
effective. Further evidence for high density zones is derived from the
[O~I] and [Ni~II] spectral lines, as well as from the kinematic 
measurements of ionic species in the nebula. 
The ionization structure of iron in Orion is modeled using the newly calculated
atomic data, showing some significant differences from 
previous models. The new model suggests a fully ionized H~II region at
densities on the order of $10^3$ cm$^{-3}$, and a dynamic partially
ionized H~II/H~I region at densities of  $10^5-10^7$ \cm3.    
Photoionization models also 
indicate that the optical [O~I] and [Fe~II] emission 
originates in high density partially ionized regions within ionization 
fronts, 
thereby confirming the general
Fe~II/O~I correlation in H~II regions determined in earlier studies.
The gas phase iron abundance in Orion is estimated from observed 
spectra, including recently observed [Fe~IV] lines. 

\end{abstract}



\section{Introduction}

The spectra, ionization structure, and the abundance of iron are
valuable indicators of the physical conditions and the chemical evolution of 
astrophysical objects. First,
owing to its  relatively high abundance
and the complex atomic structure of Fe ions, a large number 
of spectral lines are observed across most of the
electromagnetic spectrum. 
Secondly, several ionization stages of iron may be observed
from different zones within the same object. 
The low ionization stages of iron, Fe~I--IV, generally span the
physical and the ionization structure
of gaseous nebulae, 
from the cool neutral regions and 
partially ionized zones, to the fully ionized regions close to the
ionizing source. For instance, since the 
ionization potentials of Fe~I and Fe~II
are 7.6 and 16.2 eV respectively, compared with 13.6 eV for H~I, the 
Fe~II spectrum commonly traces the conditions in partially ionized 
regions and ionization fronts where Fe~II is shielded by H~I. 
Beyond 54.4 eV, the ionization potential of Fe~IV, the
photon density from the radiation field of O and B stars is too low to
effect higher ionization of iron (Osterbrock 1989). Thus, a combined
study of Fe~I--IV should provide
detailed information on the structure and physical conditions of the HII region.
Thirdly, while iron 
is primarily formed in the interstellar medium (ISM) by supernovae, 
it is a refractory element and its gas phase
abundance constrains both the chemical enrichment of the ISM and the 
formation of dust grains. 

As the 
best observed diffuse H~II region, the Orion nebula is commonly
used as a ``benchmark'' for nebular studies.
The present study aims at an understanding of some aspects of 
low-excitation photoionized nebulae through the
analysis of the ionization structure and spectra of Fe~I-IV in Orion
based on: (i) new calculations of atomic data for Fe~I-IV, 
necessary for ionization equilibrium and spectral formation modeling,     
(ii) study of ionization balance,
excitation mechanisms, and radiative transfer of the spectra of Fe ions; 
(iii) study of the
physical structure of nebulae including density and temperature variations
and kinematic effects. 

Despite the importance of low ionization stages of iron 
in laboratory and astrophysical plasmas, accurate atomic data 
for these ions were not 
available until the recent advances by the Iron Project
(hereafter IP; Hummer \etal 1993). The IP
calculations are carried using the close
coupling approximation and the R-matrix method that is presently the
most accurate method for calculating photoionization,
recombination, and excitation cross sections.
The calculations are rather involved, and computationally  intensive,
owing to the complex electron-electron correlation effects, 
the large number of coupled atomic states of iron ions,
and the complex resonance structures 
present in the cross sections. 
A brief discussion of these calculations and the atomic data 
is presented in Section 2. 

Excitation mechanisms for iron spectra are of particular
interest. The
forbidden [Fe~II] emission is known to be  exceptionally strong in most gaseous 
nebulae, such as Orion 
(e.g. Grandi 1975; Osterbrock \etal 1992), supernova remnants (e.g.
Dennefeld \& P\'equignot 1983;
Dennefeld 1986; Henry \etal 1984; Hudgins \etal 1990; Rudy \etal 1994),
Seyfert galaxies (e.g. Osterbrock \etal 1990), and
circumstellar nebulae of luminous blue variables (e.g. Stahl \& Wolf 1986;
Johnson \etal 1992). Bautista, Pradhan, \& Osterbrock (1994) found 
that the optical [Fe~II]
emission in Orion seems to be excited in 
regions with high electron densities of $\sim 10^5-10^7$ \cm3. 
Further, (Bautista \& Pradhan 1995a; BP95a hereafter) 
presented evidence that [Fe~II] emission
in Orion stems from partially ionized zones (PIZs), where hydrogen and oxygen
are mostly neutral, and collisionally excited [O~I] lines are
observed. 
At about the same time, in order to explain the anomalously intense
[Ni~II] optical emission from a variety of astrophysical objects, Lucy (1995) 
showed that the [Ni~II] emission 
in Orion and the circumstellar nebula P Cygni 
could be excited by fluorescence via the background UV continuum, and 
suggested that [Fe~II] optical emission  may be similarly affected by UV
fluorescence. This 
photo-excitation mechanism for [Ni~II] and [Fe~II]
was further examined by Bautista, Peng, \& Pradhan 
(1996; BPP hereafter).
It was concluded that owing to the differences in the atomic structure, 
fluorescence excitation is less effective for [Fe~II] than for [Ni~II] and   
appeared relatively unimportant for Orion. 
In contrast, Baldwin \etal (1996; B96 hereafter) modeled the optical [Fe~II] emission 
from Orion and found UV fluorescence to be a viable mechanism
as opposed to the high
electron densities deduced by BPP. They argued that
the UV lines of 
Fe~II that dominate the photoexcitation are optically thick 
and by including self-shielding, a process not considered in BPP, 
the observed [Fe~II] spectra could be reproduced 
at electron densities of $\sim 10^4$ \cm3.
The collisional and fluorescent excitation, 
including optical depth effects, are further investigated herein 
(Section 3.1). Our results, and those reported by B96,  are compared 
with several independent spectroscopic measurements of Orion.

In Sections 3.2 and 3.3 we investigate the excitation of [Fe~III] 
and 
[Fe~IV] lines
to study the physical conditions in the emitting region and 
the abundance of iron in Orion.
It is expected that optical [OI] emission arises from the same partially 
ionized regions as the [FeII] spectrum. In Section 4 we discuss the 
[OI] line diagnostics with presently available measurements.

The role of kinematics in the structure and physical properties of 
photoionized nebulae is also investigated.
In diffuse H~II regions, such as Orion
and M17, the ionizing radiation from hot stars 
drives dissociation (D) and ionization (I) fronts into a dense molecular cloud  
on the 
far side of the nebula. The newly ionized gas is driven 
by a strong pressure gradient, the ``champagne''
effect (e.g. Bodenheimer \etal 1979), and results in a
stratification of velocities and 
ionization states of the emitting ions in the H~II regions. 
Such a correlation  between velocities and ionization was
first observed by Kaler (1967) and is discussed in Section 5 of the 
present paper. 
In Section 6,
some discrepancies between the predictions of  current static 
photoionization models and the observed structure of nebulae are pointed out.
Photoionization models of Orion, taking into
account the high density ionization, fronts are presented together with
the predicted spectra of neutral and low ionization species, e.g. 
O~I and Fe~II.

Finally, we derive gas phase Fe/O abundance ratios in 
different emitting regions in Orion from spectroscopic data (Section 7). 
Section 8 summarizes 
the results and the conclusions.

\section{Atomic Data}

We present a review of the atomic data for 
spectroscopic analysis and radiative modeling of Fe~I--IV. This 
includes: electron-impact excitation rate coefficients 
($\Upsilon$'s), 
radiative transition probabilities ($A$-values), 
photoionization cross sections, and recombination rate coefficients. 
These data have been calculated at Ohio State in collaboration with 
S. N. Nahar and H. L. Zhang. Table 1 summarizes
the extent of the computed data employed in the 
present work. 
The new IP data
complements or supersedes the radiative data for the iron ions obtained
under the Opacity Project
(Seaton \etal 1994).

With the exception of  experimentally measured energy levels (compiled
by Sugar and Corliss 1985) and oscillator strengths for dipole allowed 
and intercombination transitions in FeII, 
all other atomic data for iron ions are computed theoretically. 
The complexity of the atomic structure of these ions
places formidable demands on large-scale computations
using the most accurate methods, and
the accuracy of some of the data may be constrained by present
computational resources. Nevertheless, by comparing several sets of
independent calculations for a given ion, and for different but related
atomic processes such as bound-bound and bound-free radiative
transitions, it is possible to derive some
conclusions about the uncertainties of the data and the reliability of
the resulting atomic models.

\subsection{Collision strengths and transition probabilities}

 The collisional-radiative models used in the present study employ the
Maxwellian averaged effective collision strengths and transition probabilities
for forbidden transitions 
obtained as described below.

\subsubsection{Fe~II}
 
 Recently, collisional data for electron impact excitation of
Fe~II were provided by the close-coupling calculations of 
Pradhan \&  
Zhang (1993) and Zhang \& Pradhan (1995a; ZP95 hereafter). 
Collision
strengths and rate coefficients were given for 10,011 transitions among
142 fine structure levels of 38 LS
terms dominated by the configurations $3d^6\ 4s,\ 3d^7,$ and $3d^6\
4p$. However,
these data sets did not
include levels with multiplicity ($2S+1$)=2 or levels dominated by the $3d^5\ 4s^2$ (e.g.
the $a\ ^6S_{5/2}$ level). In order to complement ZP95 and
to check on the accuracy of these data for optical transitions, we
carried out a new calculation (Bautista \& Pradhan 1996; hereafter, BP96) 
including the lowest 18 multiplets of Fe~II with 52 fine
structure levels and 1326 transitions,  using 
an improved wavefunction representation of the ion.
Although the BP96 collision strengths exhibit background values in very good
agreement with those of ZP95, there are differences between the two sets
owing to the resonance structures belonging to
the odd parity terms of the
$3d^6\ 4p$ configuration. 
In order to estimate the effect of the coupling to the odd parity
terms, we have carried out a new
23 state close coupling
calculation (23CC) that includes the $3d^64p\ z\ ^6D^o, z\ ^6F^o, 
z\ ^6P^o, z\ ^4F^o,$ and $ z\ ^4D^o $ terms in addition to the 18
multiplets in BP96. Some 
additional resonance
structures, absent in BP96,
enhance the Maxwellian averaged collision strengths $\Upsilon(T)$, 
resulting
in better agreement with ZP95 for the transitions common to the two
datasets.
Table 2 presents a comparison between the $\Upsilon$-values at $10^4$K from
the new 23CC calculation and the BP96 and ZP95 calculations for
transitions from the ground state $^6D_{9/2}$ with $\Upsilon > 0.1$.
Only a few such transitions show a significant difference with respect to 
BP96,  e.g.
the $^6D_{9/2} - ^6S_{5/2}$ transition
for which $\Upsilon(10^4$K) increased from 0.399 to 0.857 due to
additional resonance contributions.
The present 23CC results are well correlated with those of ZP95, particularly 
for transitions from the levels of the ground ($a\ ^6D$) and first excited
($a\ ^4F$) multiplets which dominate the collisional excitation of the ion. 
There are, however, two sets of transitions, $a\ ^6D_i-a\ ^4F_j$ and 
$a\ ^4F_i-b\ ^4F_j$,
for which the collision strengths differ by factors of two to three. 
This discrepancy suggests that there may be some
difficulties in representing the states of FeII with a $3d^7$ configuration, 
which would require a more extensive target basis set 
than used so far.
Nevertheless, the uncertainty in excitation rates for these $3d^7$ levels       
has only a small effect on the 
near-IR and optical line ratios, as shown  
in Sections 3.1 and 3.2.
For the present work we construct a 159 level system for Fe~II that
uses data for the 142 levels in ZP95, and the new results from the 
23CC calculation.

Accurate transition probabilities for forbidden transitions are difficult
to calculate because very accurate wavefunctions are needed to obtain
the weak relativistic electric quadrupole and magnetic dipole probabilities
with very small Einstein A-coefficients.
 For Fe~II such computations  are especially challenging
owing to the number of algebraic terms involved.
The first A-values for the [Fe~II] IR and optical lines were reported by 
Garstang (1962).
Nussbaumer \& Storey (1988) provided a more accurate set of A-values
for IR and near-IR transitions that result from the 16 fine structure levels of the first 4 LS
terms, $3d^64s\ a^6D,\ a^4P$ and $3d^7\ a^4F, \ a^4D$,  using the program
SUPERSTRUCTURE (Eissner \etal 1974).
Recently, Quinet \etal (1997) have reported more extensive calculations for
the [Fe~II] transitions, using SUPERSTRUCTURE and the semi-empirical 
code by Cowan (1981).
The differences between these data and those of Nussbaumer \& Storey
and Garstang's are small, a few percent for most transitions, but they 
reach up to a factor of two for some transitions.
Computational limitations 
restricted Quinet \etal from explicitly including some important
configurations that were included earlier by Nussbaumer and Storey
(1988), such as $3d^54d^2$ and $3d^54f^2$, in the configuration
interaction expansion. The effect of excluding these configurations was
estimated to be approximately 25-30\% for many transitions.

Some indication of the accuracy of the different A-values data sets may be
obtained by comparing with observations  for pairs of lines that
originate from the same upper level; then the intensity ratio 
depends only on the known energy  differences and the A-values.
In Tables 3 and 4 we compare such line ratios, for IR and optical lines
respectively, with observations of the Herbig-Haro
object, HH1. This object exhibits particularly strong [Fe~II] emission
while emission from HI and HeI are 
inhibited.
Similar comparisons have been carried out using spectra of various nebulae and
supernova remnants. 
Table 3 reveals very good agreement, within $\sim 5\%$ for most line ratios,  between the data by
Nussbaumer \& Storey and the observations, but the data
by Quinet \etal yield significant discrepancies. One
particularly interesting ratio is I(1.257\mic)/I(1.644\mic), comprising
the strongest [Fe~II] lines in the near-IR region. This ratio is
useful, for example, in the diagnosis of dust extinction 
(e.g. Dennefeld 1982, 1986;
Bautista, Pogge, \& Depoy 1995). The observed ratio of 1.34 agrees
within 2\% with the expected value from the Nussbaumer \& Storey data, while
the Quinet \etal data yield a ratio nearly 30\% lower. 
The line ratio comparison for optical lines in Table 4 shows
some differences between observations and Garstang's data,
$\sim 30\%$. The differences with respect to the data by Quinet \etal
are somewhat larger. 
 
Based on these comparisons we adopt the
data by Nussbaumer \& Storey
for the IR transitions, and Garstang's data for the rest. We
estimate that these data sets may be uncertain to the 5-10\% and
30-40\% levels respectively. More calculations
of transition probabilities for the optical [Fe~II] lines are needed.
 
\subsubsection{Fe~III}
 
Recently, Zhang \& Pradhan (1995b) and Zhang (1996) reported an
extensive set of collision strengths for transitions among 83 LS terms and 
219 fine structure levels. They investigated and found 
the relativistic effects on the collision strengths for the low-lying
forbidden transitions to be small, but the effects of resonances and
channel couplings to be considerable. They also reported good agreement
between a much smaller
7-term calculation and  earlier results by Berrington \etal
(1991). However, the full 83-term calculation yields   
resonances that 
enhance the $\Upsilon$-values by up to a factor of two. 
As part of the IP series of publications, Zhang (1996) has 
provided the complete dataset of the Maxwellian averaged
$\Upsilon$-values for the 23,871 transitions among 219 levels 
in Fe~III (Table 1).
 
Only two calculations of transition probabilities for [Fe~III] lines
have been reported. The first calculation was carried out by Garstang
(1957). More recently, Nahar \& Pradhan (1996) presented new A-values
for a larger set of forbidden transitions than the Garstang work, 
as well as for  a large number of dipole allowed transitions.
In general, the results of Nahar \& Pradhan agree within $\sim 30\%$
with those
of Garstang for the transitions common to the two datasets, with some
differences of factors of two or higher. 
The differences are sufficient to significantly affect 
several diagnostic line ratios; a comparison with available
spectroscopic observations (line ratios) shows that the
new forbidden A-values of [Fe~III] should be preferred (Nahar and
Pradhan 1996).
 
\subsubsection{Fe~IV}
 
The latest collisional calculation for Fe~IV has
been carried out by Zhang \& Pradhan (1997). This work includes transitions
among 49 LS
terms and 140 corresponding fine structure levels, 
dominated by the $3d^5$, $3d^44s$
and $3d^44p$ configurations. Zhang \& Pradhan also did a 5-term calculation
whose results compare very well, within $\sim 5\%$ or better,
with the earlier 4-term ($3d^5\ : ^6S, ^4G, ^4P$, and
$^4D$) calculation of Berrington \& Pelan (1995; some of these values 
were revised in an Erratum by Berrington \& Pelan 1997).
However, Zhang and Pradhan (1997) find that the larger 49-term 
expansion, including the couplings to the $3d^4\ 4p$ 
odd parity terms, leads to resonance
structures that enhance the $\Upsilon$-values by 20 to 50\% for many
transitions over the Berrington and Pelan data. We expect the
overall uncertainties in the Zhang and Pradhan data to be 10-30\%.
 
The only calculation of A-values for [Fe~IV] lines that has been
reported is that of Garstang (1958). Spectroscopic observations of
these lines are very scarce. Therefore, no checks on the reliability
of these data can be made. New    
calculations of forbidden transition probabilities that take into account 
important 
spin-orbit and spin-spin relativistic perturbations are needed.  

\subsection{Photoionization cross sections and recombination 
rate coefficients}

 The photoionization models depend on the accuracy and consistency
of the photoionization and the recombination data. During the past few
years the close coupling approximation 
has been extended to a unified treatment of electron-ion recombination
that includes both radiative and dielectronic recombination processes
(Nahar and Pradhan 1992;
1994a). Further, since the same target eigenfunction expansion is employed
for the photoionization and the recombination calculations, the
photoionization/recombination data are self-consistent. 
 Photoionization cross sections and the unified electron-ion
recombination rate coefficients have been computed for all iron ions of
interest herein: Fe~I--IV. 

The new photoionization cross sections
differ substantially, by up to orders of magnitude in the important
near-threshold regions, from earlier calculations by Reilman and Manson
(1979; RM hereafter) and Verner \etal (1993), both using the central field
approximation that neglects
the spin and angular multiplet atomic structure. Their results
contain
discontinuous and unphysical jumps in the photoionization cross
sections and
underestimate the Fe~I photoionization cross sections by up to 
three orders of magnitude (Bautista and Pradhan 1995b, Bautista 1997), 
the Fe~II and Fe~III cross section by up to one to two orders of magnitude 
(Nahar \& Pradhan 1994b; 
Nahar 1996a), and the Fe~IV and Fe~V cross sections by about one order of magnitude 
(Bautista \& Pradhan 1997; Bautista 1996).
The present photoionization cross sections also differ by orders of
magnitude from the calculations by Sawey \& Berrington (1992) who used 
a restricted set of eigenfunction expansions for Fe~I--IV that did not
include sufficient configuration interaction.
At high energies, however, the present photoionization cross sections are in
reasonable agreement with those obtained from the central field type 
approximations (expected to be
reliable at high energies). It is also of interest to note that
the present ground state photoionization cross section of Fe~I is in good
agreement with that computed by Kelly (1972) 
using the many-body perturbation method.  Fig. 1 shows 
the ground state photoionization cross sections 
of Fe~I--V together with several previous results. The preponderance of the
extensive resonance structures, and their effect on the effective cross
sections, is readily discernible.

Unified and total 
electron-ion recombination coefficients have been recently calculated 
for Fe~I (Nahar, Bautista, \& Pradhan 1997), Fe~II (Nahar 1997), Fe~III (Nahar 1996b), 
and Fe~IV (Nahar, Bautista, \& Pradhan; to be submitted).
The computations employ the same close coupling 
expansions and wavefunctions as used in the photoionization calculations. 
This enables, for the first time, 
complete consistency between photoionization and  
recombination data in the photoionization models.
The new recombination rates for Fe~I--III are shown in Fig. 2, and
compared with previous results by Woods \etal        
(1981). The new rates for Fe~I and Fe~II are about a factor 
of 5 higher than those by Woods \etal in the temperature range  
$10^3$ to 10$^4$K. For FeIII 
the differences are about a factor of 2 at $T=10^4$K. An interesting
feature of the new unified rates is that at high temperatures, where 
dielectronic recombination dominates the total recombination (around the
large `bump'), the present rates are substantially lower than the rates
derived from the Burgess General Formula (used by Woods \etal) that is often
employed. This large reduction is due to autoionization into the
numerous excited states of the target ion, not considered in the Burgess
formula. 

\section{Analysis of the Iron Emission Spectra}

 Spectral diagnostics of iron ions with several collisional-radiative models and
observations in different wavelength regions are described.

\subsection{The optical [Fe~II] lines}                       

In previous work BPP considered two 
different collisional-radiative (CR) models for Fe~II: a model 
including the lowest 52 even parity levels     
(doublets, quartets, and sextets) and a 142 level model including even and odd parity 
levels with multiplicities (2S+1) =  4 and 6. This larger model was 
used to investigate the fluorescence 
excitation mechanism. For the present work we have constructed 
an extended 159-level CR model for Fe~II that combines the previous 
142 quartet and sextet levels with the lowest doublet levels 
($a\ ^2G_j, a\ ^2P_j, a\ ^2H_j, a\ ^2D_j, b\ ^2P_j, b \ ^2H_j, a\ ^2F_j,$ 
and $b\ ^2G_j$)  and the previously omitted   
$4s^2\ ^6S_{5/2}$ level. 
Dipole allowed and intercombination transition probabilities, 
necessary for fluorescent excitation, are  
taken from the compilation of experimental and solar data by Giridhar \& Ferro (1995), when available, and from Nahar 
(1995) for the rest of the transitions.
An energy level diagram for Fe~II is shown in Fig. 3. Many of the 
lines observed in Orion's optical and near-IR spectra are 
also marked in this figure.

BPP discussed that, unlike the case of Ni~II (Lucy 1995) where all
levels participating in fluorescent excitation have the same spin
multiplicity (doublets),
photoexcitation of Fe~II is a relatively inefficient mechanism
and the critical densities for fluorescence of most lines
are only of the order on $10^4$ to $10^5$ \cm3 \ 
for the radiation field in Orion.
This is because the ground state of Fe~II is $^6D$,  while
most of the observed lines involve the quartet multiplets. 
Thus, photoexcitation of quartet levels must occur via intercombination 
transitions which are much weaker than the dipole transitions in low-ionization
atomic species where
relativistic spin-orbit mixing is weak.
This is the case for Fe~II, as revealed by inspection of the A-values.
We found that photoexcitation of Fe~II 
could not explain the observed relative intensities of optical [Fe~II] lines. 
Nevertheless, recently B96 and Rodr\'{\i}guez 
(1996) have suggested that [Fe~II] lines in Orion are 
dominated by the fluorescence mechanism. Therefore, further 
investigation of this issue is necessary.

In terms of the atomic data the present fluorescent excitation model should 
be quite accurate since we use experimental A-values for most dipole and 
intercombination transitions. Also, 
it was shown in BPP that the efficiency of
the photoexcitation by UV continuum radiation mechanism drops
rapidly with the energy of the odd parity levels, thus
the number of energy levels included in the present
model should be sufficient.
One aspect of the present fluorescent excitation model that was 
neglected in BPP is the effect of radiative transfer. B96
have suggested that the UV lines dominating the pumping are optically thick 
and this would affect fluorescent excitation. 
The line center optical depth for a transition $u-l$, where $u$ and 
$l$ are the upper and lower levels respectively, is given by 
\begin{equation}
d\tau_{ul}={\pi e^2\lambda f_{lu}\over mcv}(n_l-n_u g_l/g_u) N_i dr
\end{equation}
where $v$ is the linewidth ($\sim 13\times 10^5$ cm s$^{-1}$ for Orion), $g_l$ and $g_u$ are the 
statistical weights of the levels, $n_l$ and $n_u$ are the population 
fractions for the levels, and $N_i$ is the number density of the 
ion $i$ being considered (Mihalas 1978). In plane parallel geometry 
the probability that a line photon will escape from the absorption 
layer is given by
\begin{equation}
P_{esc}(\tau_{ul})= {1-e^{-\tau_{ul}} \over \tau_{ul}}.
\end{equation}
With this expression for the $P_{esc}$ one can compute the emissivity
of the lines in NLTE, including the effects of line self-absorption, by
replacing the A-values by effective
transition probabilities,
\begin{equation}
A_{ul}' = A_{ul} P_{esc}(\tau_{ul}).
\end{equation}
Here, the total $\tau_{ul}$ along the line of sight can  be evaluated 
from eqn. (1) by adopting some reasonable column density of Fe~II 
ions. This column density takes values between 0 at the near side 
of the cloud and a maximum at the far side of about 
$N_{Fe~II}\times r=N_e\times (N(Fe)/N_e)\times (N(Fe^+)/N(Fe))
\times 0.1$ pc $\simeq 8\times 10^{13}$ cm$^{-2}$, where $N_e=4000$ \cm3,
 $(N(Fe)/N(H))=10^{-5.5}$  
assuming a depletion factor for iron of ten, and $(N(Fe^+)/N(Fe))\sim 0.02$. 
For most of the present calculations we use a mean column density 
of $4\times 10^{13}$ cm$^{-2}$.
The sensitivity of the results to variations of column density 
of up to an order of magnitude were also investigated. 

To calculate theoretical emissivities and line
ratios for the ion it is necessary to simultaneously solve the 
coupled equations of radiative transfer and statistical equilibrium in
an iterative manner. 
In the present calculations we consider convergence of the level 
populations to better than 1\% for all the levels. This is attained
typically after only two or three iterations, 
since optical depth effects are small and have only marginal importance. 
The only 
transitions that can be considered as optically thick, $\tau > 1$, 
are the dipole allowed transitions that involve the ground state of the ion: $a\ ^6D_{9/2}-z\ ^6D^o_{9/2}$, 
$a\ ^6D_{9/2}-z\ ^6F^o_{11/2}$, and $a\ ^6D_{9/2}-z\ ^6P^o_{7/2}$. 
Other transitions, most of them intercombination transitions which 
dominate the radiative cascades, exhibit optical depths $\lesssim 0.2$. 

 We identify particular transitions that are relatively
insensitive to fluorescence 
and which afford reliable density diagnostics.
 To that end, we compared the population of all 159 levels 
in our Fe~II model for both the collisional and fluorescence models 
under conditions of \te=$10^4$ K and \ne=4000 \cm3, and a radiation field 
as in BPP and Lucy (1995). Then, we identified the levels that 
are least affected by 
fluorescence. As expected, among these levels are those with 
multiplicity two (e.g. $a\ ^2G_{9/2}$ which gives rise to the 7155 
and 7452 \AA\  lines), since they are not directly coupled  to 
the sextet ground state. 
Other levels nearly insensitive to fluorescence are 
$a\ ^6D_J, a\ ^4F_J, a\ ^4D_J$, and $a\ ^4P_J$, which yield  
all of the IR and the near-IR lines, as well the 8617 and 8892 \AA \ lines.       

Emissivity line ratios for lines insensitive to fluorescence are shown in 
Figs. 4(a)-(c) together with the observed ratios by OTV and 
Rodr\'{\i}guez (1996).
Several other line ratios are shown in Figs. 4(d)-(l). Here, the different 
curves represent pure collisional excitation, collisional 
and fluorescence excitation 
with and without optical depth effects for the radiation field expected 
in Orion, 
and fluorescence excitation by a radiation  
field ten times stronger than in Orion.
The observed line ratios are represented by horizontal lines. 
Also in these figures, the line ratios predicted by the three models
of B96 are shown as filled squares. 

Several conclusions can be derived from these figures. First, 
the calculated line ratio curves with and without
optical depth effects for the UV lines are nearly indistinguishable.
Therefore, optical depth effects under the nebular conditions in Orion 
have negligible effect on the fluorescent excitation of the optical [Fe~II] 
emission. 
Secondly, the present results confirm earlier works (e.g. Bautista, Pradhan, \& Osterbrock 1994 and BPP)
 showing that the optical [Fe~II]
line ratios are consistent with high densities ($10^5 - 10^7$ \cm3) 
regardless of the choice of 
collision strengths among those currently available.
This is particularly the case for line ratios unaffected by 
fluorescence, which at \ne=4000 \cm3 \  would yield
line ratios different from  the observations  by more than a factor of two. 
Among the line ratios that are affected by fluorescence, only a few 
seem to agree with the observations, while a majority 
exclude this excitation mechanism. 
This is also the case for the 
line ratios considered by Rodr\'{\i}guez (1996), e.g. 
I(7155)/I(8617) (Fig. 4(c)), I(4287)/I(8617) (Fig. 4(j)), and 
I(5262)/I(8617) (Fig. 3(b) in BPP and Fig. 4(c) in BP96). 

 For several of the line ratios (Figs. 4(e-l)) the present model agrees 
reasonably well with the
different models presented by B96. However,
their model of Orion entails a mean density for the
fully ionized zone (FIZ) of $\sim 10^4$
\cm3, which is more than twice the density (4000 \cm3; OTV) normally
derived from [O~II] and [S~II] line ratios along the
line of sight. This has the effect on the model of reducing the depth
of the FIZ and the geometrical dilution of the radiation which is
inversely proportional to the square of that depth. The higher density
in their model should lead also to an overestimation of the optical depths. 
An unexplained discrepancy between
our results and those of B96 appears in the
fluorescent pumping of the 4277 \AA
($a\ ^4F_{7/2}-a\ ^4G_{9/2}$) line.
In general, it is found that both our present fluorescent model and 
those of B96 
fail to reproduce the observed line ratios regardless of any possible
enhancement of the stellar radiation field.
 
Rodr\'{\i}guez (1996) compared observed [Fe~II] line ratios 
from 12 different positions in Orion, and from 16 positions in
6 other H~II regions, with the ratios expected 
under collisional excitation conditions. Rodr\'{\i}guez reported 
considerable scatter for the line ratios and the predictions from BPP's 
collisional model. This scatter 
may be attributed to a variety of 
causes. One is the combined uncertainties in atomic data and observations. 
We note that Rodr\'{\i}guez applied the same extinction 
correction to every spectrum;
however, the extinction in Orion is known to vary widely on small 
spatial scales (e.g. OTV; Pogge \etal 1992; Bautista \etal 1995). In 
addition, uncertainties from individual line intensity measurements themselves 
may be large and should be considered. 
One particular line that was affected
by errors in the atomic data is the 4287 \AA\  line (see previous Section)
for which Rodr\'{\i}guez found the largest discrepancies. The new
excitation rates for the $a\ ^6S_{5/2}$ upper levels of this line
enhance the predicted I(4287)/I(8217) ratio by about a factor of two. 
Another source of dispersion comes from the fact that 
a single temperature was assumed in trying to
match the observations for twelve different positions in M42 and
nine observations of other objects; but the ratios presented are
temperature sensitive, particularly at densities of $\sim 10^6$ \cm3
and higher. For instance, the I(5262)/I(8617) ratio at \ne=$10^6$ \cm3
varies by more than a factor of two between 5000 and 10000 K
(see Fig. 4(c) in BP). One more reason for the observed scatter, and 
perhaps the most important, 
is that if the PIZ is a thin transition region it is expected that the 
physical conditions vary rapidly within it, as suggested in BP95a. 
Then, a single set of \te \ and \ne \ may not 
fit all the line ratios simultaneously. 
Nevertheless, every line ratio reported by Rodr\'{\i}guez indicates 
electron densities between $10^5$ and $10^7$ \cm3, while no definitive
dependence on fluorescent excitation is found.

Additional spectroscopic evidence against fluorescent excitation of Fe~II 
in Orion is the absence of some allowed emission lines in the observed spectra. 
If the population of the levels that give rise to the forbidden lines were 
dominated by cascades from the odd parity levels, the allowed transitions 
that result from these cascades, some of which lie in the optical region, 
should be seen. That is the case, for instance, of the 5169.0 \AA 
\ ($z\ ^6P^o_{7/2} - a\ ^6S_{5/2}$)     
line that could arise due to the fluorescent excitation of the $^6S_{5/2}$ 
level, via the sequence $^6D_{9/2} - ^6P^o_{7/2}- ^6S_{5/2}$. Then,
the strength of this 
line should be directly related to the strength of the 
4287 \AA \ ($^6S_{5/2}-^6D_{9/2}$) feature and, under the conditions in Orion, 
the intensity of the 5169.0 \AA \ line should be about 70\% of that at 4287 \AA . 
Other allowed transitions of similar intensity are at 3227.73 \AA \ ($z\ ^4D^o_{7/2} 
- a\ ^4P_{5/2}$) and 3259.05 \AA \ ($y\ ^4F^o_{9/2} - b\ ^4D_{7/2}$). None 
of these lines has been observed in Orion. Moreover, recent echelle   
observations by Peimbert \etal (1996) establish an 
upper limit to $I_{Fe~II}(\lambda 5169)/I_{[Fe~II]}(\lambda 4287)$ of 
about 0.1, given by the sensitivity limit of their spectra. This 
indicates that less than 20\% of the total intensity of the [Fe~II] 4287 \AA\ 
line in Orion could be explained by fluorescent excitation by  UV continuum 
radiation. 

In contrast with Orion and other diffuse H~II regions, circumstellar nebulae 
and the subclass of bipolar planetary 
nebulae with symbiotic star cores do exhibit rich Fe~II spectra, in addition 
to the forbidden [Fe~II] lines, particularly in their core regions, e.g. Eta Carinae
(Hamann \& DePoy 1994), IRAS 17423-1755 (Riera \etal 1995), 
He2-25, Th2-B, and 19W32 (Corradi 1995), and M2-9 (Torres-Peimbert \& 
Arrieta 1996). Preliminary comparisons indicate good agreement between our 
fluorescent model and the observations of these objects (to be reported
in a later publication).

\subsection{The IR and near-IR [Fe~II] lines}

The strength of near-IR [Fe~II] lines with respect to the optical lines 
also indicates the presence of high density regions. The 12567 \AA
\ ($a\ ^6D_{9/2}- a\ ^4D_{7/2}$) line from Orion was measured by Lowe \etal (1979) 
using a large circular aperture (2' in diameter) that contains the 
region studied by OTV. More recently, we also measured this line 
(Bautista \etal 1995) centered at the same location as OTV, although 
the effective aperture was only about half of that of OTV. Because of 
the large changes in local extinction at small scales and the 
filamentary structure of the low ionization emission of Orion (e.g. Pogge
\etal 1992) near-IR to optical line ratios from independent observations 
covering unequal areas may be unreliable. 
Nevertheless, B96 used the 
intensity ratio between the 8617 \AA \ from OTV and the 12567 \AA \ line from 
Lowe \etal \ as evidence against the high density regions. This 
and other ratios of optical lines to the 12567 \AA\  line are shown 
in Figs. 5(a)-(d). Two sets of observed line ratios are given here according 
to the 12567 \AA \ intensities from Lowe \etal and Bautista \etal\  (about 
30\% greater).
While the I(12567)/I(8617) ratio seems consistent with \ne about $10^4$ 
\cm3, all other optical to near-IR line ratios yield densities greater 
than $10^5$ \cm3. As for the optical line ratios, neither the 
fluorescent models of B96 nor ours  can reproduce the 
majority of the observed line ratios. 
Future analysis of IR and near-IR [Fe~II] lines is necessary and will need 
spectroscopic observations of the nebula with sufficiently high resolution
to separate most of the [Fe~II] lines from the much stronger H~I emission.

In conclusion, while the intensity of near-IR lines with respect to the 
optical lines seems consistent with the existence of high density regions 
in the Orion nebula, some systematic differences between 
optical and IR lines may exist, as pointed out in BP95a,
and may be indicated by the I(12567)/I(8617) line ratio. Such differences are
expected if the PIZ is a thin transition region between the low density, 
fully ionized medium and the high density neutral medium. Then, the higher 
excitation optical emission would come preferentially from the highest densities 
zones, while IR lines, with much lower critical densities, may arise 
from the more extended lower density gas, as discussed in the next Section.

\subsection{Two-zone model of [Fe~II] emission}

The analysis of optical [Fe~II] emission from Orion
led to the discovery of high density 
(\ne =$10^5 - 10^7$ \cm3) PIZ's, but the approximation of a single
temperature and density in the emitting region may be responsible for some of 
the observed dispersion between different line ratios. 
Moreover, it is possible that some 
fraction of the lower excitation lines may originate from the FIZ. 
A more realistic model of the PIZ 
should take the small scale variations in physical conditions into account.
We construct a two-zone model for the [Fe~II] emitting 
region to illustrate how its inhomogeneity could lead to the observed \ne       
dispersion.

Assuming the dominant [Fe~II] optical emission from 
the PIZ with (\ne, \te)=($10^6$ \cm3; 10,000 K), 50\% of the gas 
ionized, and nearly all iron as Fe$^+$, 
the intensity of the 4277 \AA \ line, for instance, with 
respect to H$\beta$, may be expressed as
\begin{equation}
{I_{[Fe~II]}(4277) \over H\beta} = {j(4277) \over j(H\beta)} 
 {N_{PIZ}(Fe^+)\over N_{FIZ}(H^+)} {l_{PIZ}\over l_{FIZ}},
\end{equation}
where $j$ is the absolute emissivity of the line per ion, $N_{PIZ}$ and $N_{FIZ}$ 
indicate densities in the PIZ and FIZ 
respectively, and $l_{PIZ}/l_{FIZ}$ is the mean ratio of the column lengths 
of the two media. From the measured intensity 
of the 4277 \AA \ line with respect to 
H${\beta}$, equal to  4.3$\times 10^{-4}$ (OTV), and the FIZ density 
of 4000 \cm3, the ratio of the depths of the PIZ and FIZ is of the order  
\begin{equation}
{l_{PIZ}\over l_{FIZ}} = 10^{-6} - 10^{-7}.
\end{equation}

Thus, the PIZ appears to be a  very thin region compared with the extent of   
both the FIZ and the ionization-dissociation front. 
As a  consequence, 
a very small fraction of Fe$^+$ in the FIZ could be sufficient to 
dominate the emission of lines with low critical densities.
For instance, for the 12567 \AA \ line, with a critical density 
of the order of 10$^4$ \cm3, an increase in \ne from 4000 to $10^6$ 
\cm3 enhances its emissivity by a factor of 
only 10.4, much less than the factor of $\sim 250$ for the optical lines. Therefore, even if 
only $\sim 2\%$ of the iron in the FIZ is in Fe$^+$, one can express the 
intensity of this line by
\begin{equation}
I(12567)= (j(12567)N(Fe^+)l)_{FIZ} + (j(12567)N(Fe^+)l)_{PIZ},
\end{equation}
to obtain $ I(12567)= (j(12567)N(Fe^+)l)_{FIZ}\times (1 + 0.1)$, (i.e.
with the contribution of the PIZ about one tenth that of the FIZ). 
The 12567 \AA \ line, and the IR and near-IR 
lines with similarly low critical densities, 
should originate preferentially from the FIZ, unlike the optical lines. 
Two predictions may be made from such a scenario: (1) 
line density diagnostics with IR lines may behave differently than those 
with optical lines and would be consistent with the conditions of the 
FIZ, e.g. (\ne, \te)$\approx$ (4000 \cm3, 9000 K); (2) there should be clear 
kinematic differences between the IR and optical lines, i.e. the  
velocities measured from the IR [Fe~II] lines should be similar to those of 
the nebular [S~II] emission, 
unlike the velocities from optical [Fe~II] lines which are 
close to those of [O~I] at the ionization front 
(see Section 4). 

Table 5 shows the relative intensities of the optical and near-IR lines 
calculated 
with a model that combines a FIZ and a thin high density PIZ. The conditions 
in the FIZ are:  (\te, \ne)=(9000 K, 4000 \cm3), a 
thickness of 0.13 pc, and Fe$^+$ ionization fraction of 2\%; and 
the conditions in the PIZ are: (\te, \ne) = (10000 K, 2$\times 10^6$ \cm3), 
thickness of 3$\times 10^{-8}$ pc, and Fe$^+$ ionization fraction of 80\%. 
Fluorescent excitation for conditions similar to those of Orion is also 
included. This mechanism may only affect the emission from the FIZ 
where the electron density is lower than the critical density for 
fluorescence. The present results are also compared with observations 
of Orion by OTV, and with the models by B96.
The percentage contribution of the PIZ to the total intensity of 
each line for models (I) and (II) are indicated within brackets.
Also, the mean dispersion between OTV's observations and each of the 
models is given at the bottom of the table for the lines in common 
with B96.

Table 5 shows that the observed optical and near-IR spectra can be reasonably 
well explained by this simple two-zone model, while serious discrepancies 
exist when the contribution from the PIZ is neglected. For instance, it can be 
seen that the observed intensity of the $\lambda$5262
line with respect to the $\lambda$8617 differs by factors of two to three 
from theoretical models that neglect the PIZ (present model III and 
models A, B, C of B96). This ratio is one of the 
ratios considered by Rodr\'{\i}guez (1996). On the other hand, by including 
the contribution of a thin high density PIZ (present models I and II) 
good agreement is found for most lines including the near-IR lines.
The mean dispersions ($\sigma$) between observations and the various 
models are also indicated in Table 5. This also shows that the models presented here are
in much better agreement with the observations than those of B96
et al. Moreover, model (II) of Table 5 that neglects fluorescent excitation 
in the FIZ seems to be better than model (I). This
may suggest that the UV continuum flux considered in (I), as suggested by Lucy
(1995), is overestimated. On the other hand, model (III) and models (A), (B),
and (C) of B96, which neglect the contributions from the
high density zone differ from the observed line ratios by 53-80\%.
Notice also that in models (I) and (II) which best fit the 
observations the contributions	of the PIZ dominate the total intensity of the 
optical lines, particularly in model (II) where the PIZ 
is responsible for 70-80\% of the optical emission.
A better representation
of the [Fe~II] nebular emission would require radiative and hydrodynamic 
modeling of the 
structure of the PIZ, which exceeds the capabilities of current photoionization 
modeling codes. 

\subsection{The [Fe~III] lines}

Forbidden [Fe~III] lines are expected to be collisionally excited. Fluorescent 
pumping of these lines by continuum radiation is unlikely because of the 
large energy difference (more than 1 Ry) between the $^5D$ ground state of the ion and the 
first odd parity levels. Moreover, the stellar continuum radiation in this 
energy range is absorbed for the most part by hydrogen.

We use a 34-level CR model for Fe~III with collision strengths from 
Zhang (1996) and transition probabilities from Nahar \& Pradhan (1996). 
An energy level diagram for the Fe~III system is shown in Fig. 6, 
where the most important lines under nebular conditions in the optical and near-IR regions 
are indicated.  
One interesting characteristic of the Fe~III system is that the near-IR 
emission originates from higher excitation levels ($^3G_J$) than the 
optical lines (levels $^3F_J,\ ^3H_J$, and $^3P_J$), which explains 
why the near-IR [Fe~III] emission is usually very weak in gaseous nebulae.
Also, the maximum energy difference between the levels 
that give rise to the optical lines is only about 0.02 Ry ($\sim $ 
3000 K), which makes the relative line intensities of these lines 
insensitive to small temperature variations. 

Selected line ratios among optical and near-IR lines are shown in Figs. 
7(a)-(j) at  \te = 9000 K. 
Optical observations of Orion are from OTV, 
Greve \etal (1994), and Rodr\'{\i}guez (1996)
and near-IR observations are from DePoy \& Pogge (1994) and
Bautista \etal (1995). The observation by Greve \etal 
and Rodr\'{\i}guez 
include several different positions; here we present the entire 
ranges of their measured line ratios.
The observed [Fe~III] line ratios in Orion agree well with the diagnostics 
using lines of other species, such as [S~II], that indicate 
\ne of a few times $\times 10^3$ \cm3. 
The observations by Greve \etal\  indicate densities between 
a few times $10^3$ to almost $10^4$ \cm3, with the lowest \ne near 
the Trapezium and highest towards the edges of the nebula.
The observations of Greve \etal also reveal variations in the intensities 
of the [Fe~III] lines with respect to H$\beta$ along 
six consecutive positions 
on a North-to-South line centered at $\sim20^\prime$ W of $\Theta ^1$ Ori A.
Here, the 
intensity of the [Fe~III] emission seems to have a minimum near the      
Trapezium region and increases toward the edges of the nebula, 
particularly in towards the North. This behavior seems correlated 
with the [N~II] and [O~II] emission, but it is    
anti-correlated with the intensity of the He~I, [O~III], [S~III], 
and [Ne~III] lines. These correlations between the [Fe~III] emission 
and the low ionization species, and the anti-correlation with higher ionization 
ions, are consistent with a drop in the ionization of the plasma towards 
the edges of the nebula. 

\subsection{The [Fe~IV] lines}

We use a
33-level CR model of Fe~IV that includes the collisional rate
coefficients from 
Zhang \& Pradhan (1997) and A-values from Garstang (1958). An energy 
level diagram for the Fe~IV system is shown in Fig. 
8, which illustrates the lowest metastable levels 
$^4G_J,\ ^4P_J$, and $^4D_J$ that give 
rise to lines in the UV. Emission lines in the optical from 
higher excitation levels, about 0.5 Ry ($\sim 6.8$ eV) above the ground state,
are expected to be weak unless relatively high densities 
and/or temperatures are present. Fluorescent 
excitation by stellar continuum seems unlikely, 
as photons with energies higher than 1.7 Ry ($\sim$23 eV) are required 
to excite the lowest odd parity levels. 
We identify the strongest [Fe~IV] transitions for conditions of
\ne = 4000 \cm3 and \te = 9000 K. These UV transitions,
in decreasing order of intensity, are:
2835.7 \AA \ ($^6S_{5/2} - ^4P_{5/2}$),
2829.4 \AA \ ($^6S_{5/2} - ^4P_{3/2}$), 2567.6 \AA \ ($^6S_{5/2} -
^4D_{5/2}$),
2567.4 \AA \ ($^6S_{5/2} - ^4D_{3/2}$), and 3101.7 \AA \ ($^6S_{5/2} -
^4G_{11/2}$). The 3101.7 \AA  \ line stems from the
first excited metastable level $^4G_{11/2}$,  with a
transition probability to the ground level of only 
$\sim 10^{-5}$ s$^{-1}$; yet, it may be observable and valuable
in understanding the excitation mechanisms for Fe~IV.

Recently, the detection of the [Fe~IV] 2835.7 \AA\  line in Orion was
reported by Rubin \etal (1997) using the GHRS/HST spectrograph. 
Based on the observed 2835.7 \AA\  line
and their models, Rubin \etal \ derived the Fe/H abundance ratio  in
Orion to be lower than the solar by a factor between 70 and 200,
much lower than any previous estimate from [FeIII] or [FeII] lines. 
A reexamination of this result is presented in Section 5.

In the optical region the strongest [Fe~IV] features are those at 
4906.6 \AA ($^4G_{11/2} - ^4F_{9/2}$), 4900.0 \AA ($^4G_{9/2} 
- ^4F_{7/2}$), 4903.1 \AA ($^4G_{7/2} - ^4F_{7/2}$), 4198.2 \AA
($^4G_{11/2} - ^2H_{9/2}$), and 4152.3 \AA 
($^4G_{9/2} - ^2H_{11/2}$). However,  the strength of the 
4906.6 \AA\  line with respect 
to the 2835.7 UV feature is only about 0.014. Similarly, the strength of the 
4906.6 \AA \ line with respect to the nearby [Fe~III] 4881 \AA \ line 
is about 0.012$\times N(Fe^{3+})/N(Fe^{2+})$. These [Fe~IV] features may be 
detectable with modern instruments.  
Some potentially useful 
\ne\  and \te\  sensitive UV and optical line ratios are 
shown in Figs. 9(a)-(h).

Surprisingly, strong optical [Fe~IV] emission is seen in   
planetary nebulae with symbiotic cores like M2-9 (Balick 1989; 
Torres-Peimbert \&
Arrieta 1996). The reason for this seems to be the high \ne ($\sim 10^7$ \cm3)
in the nebular core. 
Fig. 10 compares the density diagnostic results from 
[Fe~IV] and [O~III] lines as measured by Torres-Peimbert \& Arrieta. 
The very good agreement between both diagnostics provides observational 
support for the accuracy of the atomic data and the present excitation 
model for Fe~IV.

\section{[O~I] Diagnostics}

In the optical region of the spectrum there are three collisionally
excited [O~I] lines sometimes detected in H~II regions; these are
due to the transitions $^1D_2 - ^3P_1, ^1D_2 - ^3P_2,$ and $^1S_0 - ^1D_2$
at $\lambda\lambda$ 6363, 6300, and 5577 respectively. All three lines
were reported by
OTV, and Baldwin \etal (1991) at twenty one
different positions in Orion. These observations were used by BP95a
to diagnose the physical conditions in the [OI] emitting regions, 
which were found to be in good agreement to those for optical [FeII].
BP95a also recognized that the [OI] observations may suffer to varying 
degrees from night sky contamination, so an uncertainty of about a 
factor of two on the line ratios was assumed.   

Recently, B96 carried out observations of the [O~I]
emission from Orion using the {\it Faint Object Spectrograph} (FOS)
on the {\it Hubble Space Telescope} (HST) and Cassegrain echelle
spectrograph on the 4~m telescope of the {\it Cerro Tololo Inter-American
Observatory} (CTIO). B96 report no detection of the
[O~I] 5577 \AA\  line, for a position in
Orion quite distant from that of OTV. B96 suggest that previous
measurements of the [O~I] 5577 \AA line were contaminated by telluric
emission and did not represent actual emission from Orion. Their observations 
establish lower limits to the 
I(6300+6363)/I(5577) ratio that differ 
by up to nearly a factor of four from the measurements of OTV and Baldwin \etal (1991). The HST 
and CTIO measurements of B96
establish upper limits of $10^6$ and $2\times 10^5$ \cm3 to the averaged electron density of the emitting
region assuming \te = $10^4$ K.
If temperatures of 9000 or 8000 K were adopted instead, the \ne \ upper
bounds would be $9\times 10^5$\cm3 \  (HST), $3\times 10^5$\cm3 \  (CTIO),
and $2\times 10^7$\cm3 \  (HST), $5\times 10^6$\cm3 \ (CTIO) respectively.
This is illustrated in Fig. 11. 

It is important to note that the temperature diagnostic from [NII] lines 
given by B96 is inapplicable to the [OI] region because [NII] and [OI] 
are emited by 
different regions; NII requires photon energies of about 14.5 eV 
which is higher than the formation energy of OII (see next section).
Reliable observations of optical [OI] and [FeII] lines from the same position 
in the nebula would be valuable in constraining the mean 
temperature of the PIZ.

\section{Kinematic Analysis of the Orion Nebula}

Although photoionization models generally assume static conditions,
the differential expansion of nebulae is known. For example,
Kaler (1967) presented radial velocities for a variety of ions taken from 
spectroscopic observations of Orion 
by Kaler, Aller, \& Bowen (1965). This study 
includes the expansion velocity of Fe$^+$ obtained from [Fe~II] lines. 
Fehrenbach (1977) reported additional measurements 
of three different positions in Orion 
which included the velocities from forbidden emission of Fe~II, Ni~II, and 
S~II. It was pointed out in this work that there is a 
large separation in expansion 
velocity between these low ionization species and the higher ionization 
ions, e.g. H$^+$ and O$^{2+}$. The most recent kinematic studies of Orion  
have been presented in several papers by Casta\~neda, O'Dell, and Wen. 
In particular, O'Dell \& Wen (1992) measure the expansion velocities 
with the [O~I] $\lambda 6300$ \AA \ line for several arcminutes across the core 
of the nebula.
Fig. 12 shows the measured expansion velocities of various ions from Kaler (1967), 
Fehrenbach (1977), and O'Dell \& Wen (1992) against the energy necessary 
to form them. 
This figure is adapted from plots previously presented by Kaler (1967) and 
Balick, Gammon, \& Hjellming (1974). 
The ordinate on the right 
represents the observed velocities in the heliocentric system, 
and the ordinate on the left gives the velocities with respect to 
the molecules in the OMC-1 cloud core (+27 km s$^{-1}$). 
The data by Fehrenbach correspond to the position 
GA417 in Table 3 of his paper. One difficulty with these data is that 
only statistical dispersions for the mean velocities are given, 
without including the instrumental errors which dominate when the 
number of lines observed is small. We have estimated 
error bars for these measurements by taking the ratio of the 
instrumental error ($\sim$ 3 km s$^{-1}$) divided by the square root of the 
number of lines measured. 
The data from O'Dell \& Wen correspond 
to that given in Table 5 of their paper which includes measurements 
by O'Dell \etal (1991) and Casta\~neda (1988).

Figure 12 shows a strong dependence
of the velocities of the ions on their formation energies. 
In particular, there is a sharp division in velocity between ions 
that require photon energies greater than 
13.6 eV (1 Ry), indicated by the vertical dashed line, and neutrals and
ions with
lower first ionization potentials, such as O$^0$, Fe$^+$, and Ni$^+$. 
This velocity 
stratification in Orion was previously pointed out by several authors, 
e.g. Kaler (1967) and Balick \etal (1974).
It is also clear from the figure that forbidden emission from O~I, Fe~II, 
and Ni~II should stem mostly from the same PIZ
at the ionization front, as predicted by 
photoionization models (B96) and seen from the [O~I]/
[Fe~II] and the [Ni~II]/[Fe~II] correlations (BP95; 
BPP). The velocities associated with [S~II] emission, on the other hand, 
lie in between those of the ionization front and the fully ionized zone. 
This is because, although S$^+$ requires only 10.4 eV to be created, 
it can survive to photons up to 23.3 eV. By 
contrast, Fe$^+$ ionizes to Fe$^{++}$ at 16.2 eV and, in addition, Fe$^+$ 
ionizes to Fe$^{++}$ by charge exchange with H$^+$ (Neufeld \& Dalgarno 
1987). 
Furthermore, spectroscopic diagnostics from [S~II] should 
mostly sample the region in the FIZ right behind the ionization front, and may not 
agree with the results from [Fe~II] lines.

Fig. 12 also shows that the velocities of species in the PIZ are 
similar to those of the photodissociation region (PDR) and the 
molecular cloud. This has implications for constraining 
the gas densities in that zone. 
The continuity equation
for the flow co-moving with the ionization front implies
that any two points along the gas flow with 
similar bulk velocities should have about the same density, as in the 
case of the PIZ, the PDR, and molecular cloud. From the mass conservation equation in the frame of reference comoving with the ionization front $\rho v =
constant$. Then, one 
can establish limits on the density of the PIZ by estimating
the velocity of the 
ionization front from known densities and velocities of the 
PDR and, for instance, the [O~II] emitting zone. Thus,
\begin{equation}
(10^5 \mathrm{cm}^{-3})\times (26.1 \mathrm{km s}^{-1} - u_s) = (4000 \mathrm{cm}^{-3})\times  
(13.8 \mathrm{km s}^{-1} - u_s)
\end{equation}
yields a heliocentric velocity for the ionization front of about 26.6 km s$^{-1}$.   
>From this we have
\begin{equation}
\rho_{PIZ} = 10^5 \mathrm{cm}^{-3}\times {{(26.1 \mathrm{km s}^{-1} - 26.6 \mathrm{km s}^{-1})}
\over {((25.8\pm 0.6) \mathrm{km s}^{-1} - 26.4 \mathrm{km s}^{-1})}} \ge 4\times 10^4 \mathrm{cm}^{-3}, 
\end{equation}
where $10^5$ \cm3 \ is the estimated density of the PDR (Tielens \& 
Hollenbach 1985) and 
26.1 km s$^{-1}$ is the flux averaged velocity of the PDR measured from
radio CII emission (Goudis 1982).
Therefore, the averaged density of the [O~I], [Fe~II], [Ni~II] emitting zone should be at least 
ten times greater than that at the [O~II] region (FIZ). No upper limit 
on the density of the PIZ can be obtained given the present uncertainties 
in the expansion velocities.

Apart from Orion, other H~II regions (M43, M8, M16, M17, M20, and NGC7635 
observed by Rodr\'{\i}guez 1996) also show high density PIZ's. 
It is, then, natural to 
ask: How do warm, high density PIZ's form? The answer may 
be found 
in hydrodynamic models of ionization fronts, such as those of Tenorio-Tagle 
(1977), Bedijn \& Tenorio-Tagle (1981) and Garc\'{\i}a-Segura \& Franco (1996). 
Tenorio-Tagle and Bedijn \& Tenorio-Tagle showed that I-fronts 
propagating into dense clouds are neither smooth nor in pressure 
equilibrium with the ambient gas. Rather, the I-front and the leading shock 
can give rise to highly compressed regions due to the so-called 
`rocket-effect'. The density and pressure of these regions can be orders 
of magnitude above those of the undisturbed gas. Moreover, these regions 
dissipate and reform as they become fully ionized and the shock travels 
into the molecular cloud. In their recent three-dimensional 
simulation of the forming H~II regions, Garc\'{\i}a-Segura \& Franco
(1996) also 
found that high density regions should form within I-fronts with
a shell-like structure at the interface between the H~II region and the 
dense molecular cloud. The densities of these shells may lead to
instabilities, yielding `elephant trunks' and rim-like structures.

\section{Photoionization Modeling of Orion}

The new atomic data for Fe~I--V enable
an accurate calculation of the ionization 
structure of Fe in low ionization nebulae. 
Iron has only a minor effect on the cooling of the
fully ionized region, thus changes in the ionization of iron should not
alter the conditions in the nebula.
Therefore, one can compute the temperature, electron
density, and ionizing radiation flux at every point in the nebula
using presently available photoionization modeling codes, and then use
these results to calculate the ionization-recombination balance of 
iron separately.

\subsection{Fe Ionization Balance} 
 
We use the computer code CLOUDY (Ferland 1993) for
photoionization modeling, but incorporate the new atomic data for the
Fe ions (Table 1).
Stellar continuum radiation, dust content, and turbulence velocities were
similar to
those in Baldwin \etal (1991). We assume mean chemical abundances for the Orion
nebula as in Ferland (1993), which are
based on results from Baldwin \etal, OTV, 
and Rubin \etal (1991a).
We further assume constant 
total (gas and radiation) pressure and
a mean density throughout the cloud of $\sim 10^4$
cm$^{-3}$.
This model allows us to study the effects that the new atomic data have
on the calculated ionization structure of iron in the nebula.
However, as shown later, this model seems
unable to reproduce the observed spectra of high and low ionization species
simultaneously with the observed depth of the ionized region.
 
In Fig. 13 we show the physical conditions (\nd, \te, \ne )
in the cloud, as obtained from the model. 
In Fig. 14 we present examples of the ionizing radiation flux at
two different zones in the cloud (a) at the near side to the ionizing star 
($\Theta ^1$ Ori C, spectral type O7; Conti \& Alschuler 1971)
and (b) near the ionization front. 
The ionization thresholds of Fe~I--III marked in Fig. 14 reveal the
correlation with the photoionization cross sections
in Fig. 1, near the ionization thresholds as they affect 
the total photoionization rate.
For instance, in the photoionization of Fe~I the ionizing flux per
photon energy unit is maximum between its ionization threshold 
$\sim 0.55$ Ry  (7.5 eV) and 1 Ry (13.6 eV), beyond which most of the
photons are absorbed by H~I ionization. This is
the same energy interval for which the previous photoionization data
(RM, Verner \etal 1993)
underestimates the Fe~I cross section by up to three orders of magnitude
(Fig. 1).
A similar situation applies to Fe~II whose ionization potential (1.18 Ry) 
lies below that of neutral He (1.81 Ry). As one goes deeper in the 
cloud, fewer photons with energies greater than 1.81 Ry are available and 
the photoionization of Fe~II is dominated by the flux in the near 
threshold region, where it is heavily attenuated by extensive 
resonance structures (Fig. 1).     
 
We calculate the photoionization rate of Fe~I--IV by
integrating the new photoionization cross sections over the ionizing flux
at every point in the nebula. At high photon energies where the
R-matrix data is not available, we use the central field data
by RM which should be reasonably reliable at high energies, as seen by the
relatively small differences with the close-coupling results (Fig. 1).
Photoionization from excited states of Fe ions was found to be negligible
and was not considered.
 
 Fig. 15 shows our results for the photoionization rates as a function
of distance from the ionizing star ($\Theta ^1$ Ori C), 
compared with those obtained using the cross sections of RM.
It is seen that the ionization rates for Fe~I--III calculated
with the new data increase by nearly a factor of two for Fe~I and Fe~II,
and about a factor of five for Fe~III
with respect to
those found using cross sections of RM. Near the ionization front the
photoionization rates using the new data are about an order of magnitude 
larger than those using RM.
This is because deeper in the cloud the
ionization becomes dominated by photons below the ionization thresholds
of H~I for Fe~I, and of He~I for Fe~II, where the discrepancies between the
two sets of cross sections are greater. 
We emphasized that
the resonance structures are physical features and their effect on the
photoionization cross section  should be taken into account,
and that there can be a significant loss of accuracy in using fits of
only the background cross sections.
 
The effect of charge exchange on Fe ions also needs to be considered,
particularly
\begin{equation}
Fe^{++} + H \rightleftharpoons Fe^+ + H^+ .
\end{equation}
The rate coefficients for these reactions were calculated by Neufeld
and Dalgarno (1987) using the Landau-Zener approximation and might be
highly uncertain for complex atomic systems such as the Fe ions. 
This is therefore the main source of uncertainty
for the Fe~II/Fe~III ionization balance as far as the atomic data is
concerned. For example, changes in the rates by the estimated 
uncertainty of a factor of three
varies the ionic fractions of Fe~I and Fe~II by up to 20\% (Fe~III and
Fe~IV remain almost unaffected). However, the
uncertainties  in the rates could be much higher than a factor of three, and 
new calculations or experimental measurements are needed.
Charge exchange ionization  and recombination involving
Fe$^+$ and Fe$^{++}$ are very important in limiting the physical
extent of the FeII emitting region.
 
Fig. 16 shows our results for the ionization fractions of Fe~I-IV
in the nebula, compared with those with 
earlier atomic data in CLOUDY,
which uses the RM
cross sections extrapolated to the
ionization threshold. For Fe~III for example,
CLOUDY includes a value for  the cross section at 2.2 Ry
of 8.8 Mb which is about four times higher than the background cross
section from the new data (Nahar 1996a). This overestimation of the cross
section compensates in part for the missing contribution of the resonance
structures. In addition, the code uses recombination coefficients 
from Woods \etal (1981) which are about a factor of two too low at \te around 
$10^4$ K. Therefore the agreement between the CLOUDY
predictions and the present results is somewhat fortuitous.

\subsection{Modeling of Fe in Orion}

Apart from the atomic data,
the main source of uncertainty in calculating the ionization structure of Fe in 
nebulae is the assumed structure 
for the cloud. Different assumptions about the radial density dependence
(constant, exponential, or power law), or 
constant thermal and/or radiative pressure, constant temperature, etc. 
result in significant differences in the ionization structure of Fe 
and other ions. For instance, Baldwin \etal (1991) assumed 
a mean gas density for Orion of $\sim 10^4$ \cm3, and constant 
gas pressure, and obtained Fe ionic fractions 
averaged over line-of-sight of 
(Fe$^+$/Fe$^{2+}$/Fe$^{3+}$) = (0.01/0.24/0.74). On the other hand, 
Rubin \etal (1991b) used an exponential density profile, as a function
of radius, up to a maximum of 5000 \cm3 and a ``plateau'' beyond, 
to obtain
(0.05/0.41/0.53). Some differences between these two models for the 
ionic fractions of other 
elements are also present, e.g. He. 
One might therefore expect an uncertainty of about a factor of five in 
an iron abundance estimate based on [Fe~II] lines and  
ionization corrections from photoionization models. 
If [Fe~III] lines are used instead, the uncertainty would be  about 
a factor of two. Rubin \etal (1997) have recently estimated 
the iron abundance in Orion from  [Fe~IV] lines and obtained  
values that differ by nearly a factor of three from those of  
Baldwin \etal and Rubin \etal (1991b). 

Another difficulty with modeling the ionization structure of iron, 
and particularly the ionic fraction of Fe$^+$, is the inadequacy of the
physics of
ionization fronts in photoionization models that
assume static conditions everywhere in the cloud, with the ionization           
front at a distance where all 
ionizing photons have been absorbed. However, real photoionization 
fronts are highly dynamic and 
more realistic models of `blister' 
H~II regions, like Orion, should consider the effect of ionization 
fronts where enough ionizing photons are available
to photoionize new neutral material.
One should also take into 
account the radiation energy that accelerates the gas 
away from the ionization front into the fully ionized zone.
This difficulty with the thickness of the cloud and the location of 
the ionization front is readily noticed by 
comparing the Rubin \etal (1991a) and the Baldwin \etal models. The model 
of Rubin \etal 
uses an exponentially increasing gas density with a peak value of 5000 \cm3 
and predicts a distance to the He 
ionization front of 0.277 pc. By contrast, the estimated mean thickness of 
the emitting region from the surface brightness of the H${\alpha}$ emission
is only 0.13 pc (Wen \& O'Dell 1995). Perhaps, the extent of the ionized 
region in the Rubin \etal \ model could be reduced by decreasing the number 
of ionizing photons in the model ({\it Q}), but this would conflict with  
direct measurements of {\it Q}-values from radio continuum flux density. 
The Baldwin \etal model, on the other hand, gives a thickness for the 
ionized region of about 0.07 pc, closer to observations, but the
adopted density is  $\sim 10^4$ \cm3. Such a 
density is considerably higher than what is observed 
over most of the nebula except for a region immediately 
south-southwest of the Trapezium. If a lower density had been adopted, 
like that in Rubin \etal, a much thicker ionized zone would have been 
obtained. 

In modeling the ionization structure of Orion we constructed 
a simple model of the nebula that uses a constant density of 4000 \cm3, 
based on long established [S~II] and [O~II] line diagnostics,
up to a maximum distance from the star where a high density ionization front is encountered. 
The position of the front is optimized to 
match the observed relative intensity of the [O~I] 6300 \AA \ line with 
respect to H$\beta$ without significantly affecting the intensities of [O~II] 
and [O~III] lines.            
Two models were calculated for peak particle densities 
of the ionization front of $10^6$ \cm3 and $10^7$ \cm3. Other conditions
such as
turbulent velocity, stellar temperature, and number of ionizing photons 
are as in Baldwin \etal (1991). 
There is no explicit control over the 
electron density, temperature, and depth of the PIZ, and 
they are calculated in the model according to photoionization-recombination 
equilibrium which depends only on the particle density and position of the 
ionization front. 
Table 5 shows the results, together with
three different sets of observations, columns 3-5, 
from OTV, B96, and Greve \etal (1994).
The results reported by B96 are shown in columns 
6-8. The results for five different models calculated here, (I) to (V), are 
given in columns 9-13.

Inspection of Table 5 reveals the difficulty in reproducing the observed 
spectrum with current photoionization models and, in particular, the 
problems in trying to simultaneously match all of the observed ionization 
stages of a given element, e.g. oxygen. Model (I) in the present 
work was calculated assuming constant gas pressure conditions with 
a mean density of $\sim10^4$ \cm3,
as in B96. The models systematically overestimate the intensity of   
[O~II] and [O~III] from a few percent to over a factor 
of two. On the other hand the [O~I] emission is underestimated by 
up to a factor of three. Models (II) and (III) were calculated 
with constant gas pressure conditions but with a mean 
density of $\sim4000$ \cm3 determined by the spectroscopic diagnostics. 
The difference between models (II) and (III) is that the former 
uses the LTE stellar continuum flux from Kurucz (1979), while 
model (III) uses a NLTE stellar continuum from Sellmaier \etal (1996). 
Clearly, adopting a lower density for the cloud and the use of accurate 
NLTE stellar continuum fluxes improves the results for [O~II] and 
[O~III] emission with respect to observations. However, the discrepancies 
for the [O~I] lines increase to about a factor of five or more. 
This is because lowering the mean density increases
the ionization, reducing the fractions of neutrals. 
The same conditions as in model (III) were used for (IV) and (V), 
together with high density ionization fronts in the PIZ.
Model (IV) uses a front with peak density $10^6$ \cm3 at a depth in the 
cloud of $6.5\times 10^{17}$ cm (= 0.21 pc), and  model (V) has a 
front with peak density $10^7$ \cm3 at a depth of $6.9\times 10^{17}$ cm (=0.22 pc).
[O~II] and [O~III] lines in both of these models remain nearly unaltered 
with respect to model (IV) and are in reasonable agreement with observations. 
However, the [O~I] lines are considerably enhanced, 
close to the observed levels, as a result of the contribution 
from the PIZ. 
In these models the mean temperatures and hydrogen ionization fractions,
N(H$^+$)/N(H), are 8700 K 
and 0.49 respectively for model (IV), and 8200 K and 0.31 for model (V).

Table 6 clearly shows the effect of the assumed density structure of 
the cloud on the predicted spectrum from photoionization modeling. The table 
also shows that there is a significant contribution to the emission from 
neutral and low ionization species from the high density PIZ. It must be  
pointed out, however, that the models presented in Table 6 are still
illustrative, and accurate 
modeling of ionization fronts requires a radiative-hydrodynamic 
treatment. 

\section{The Iron Abundance in Orion}

There are two basic approaches typically used to estimate gas phase 
abundances    
in gaseous nebulae. The first one consists of estimating 
the ionic abundances, normally relative to hydrogen, directly
from spectra assuming mean density and temperature 
derived from line ratio diagnostics. 
Then, the abundances of all observed ions of the same element are added 
to yield the total gas phase abundance of the element.
Few prior assumptions need to be made about the 
structure of the cloud, but the method has the disadvantage that it 
neglects any temperature and density variations
along the line of sight, as predicted 
from photoionization models (e.g. Fig. 13). Additional temperature 
fluctuations, different from photoionization models,
have also been studied by Peimbert (1967; 1995, and references therein).
The second approach consists of photoionization modeling
to reproduce the conditions in the cloud, and the abundance 
of elements adjusted to match the observed spectrum. The 
results
depend on the initial assumptions about the structure of the nebulae 
and may be subject to uncertainties in the model. 
Relative ionic abundances should be most accurate when calculated for ions 
that co-exist and whose lines are produced 
by the same mechanism, e.g. collisional excitation or recombination. 
Lines excited by fluorescence are difficult to interpret as they generally
involve a large number of levels and depend on the nebular 
photoexcitation emission. 

For Orion we calculate the 
abundances of Fe$^+$, Fe$^{2+}$, and Fe$^{3+}$ relative to 
O$^0$, O$^+$, and O$^{2+}$ respectively, and derive the 
total N(Fe)/N(O) from each of the ionic ratios using calculated 
ionic fractions.    
The abundance ratio N(Fe$^{i+}$)/N(O$^{j+}$) can be obtained from
the observed line intensities 'I', and the calculated 
emissivities 'j', as
\begin{equation}
{N(Fe^{i+})\over N(O^{j+})} = {I_{Fe^{i+}} \over I_{O^{j+}}}\times
{j_{O^{j+}}\over j_{Fe^{i+}}}. 
\end{equation}
The total N(Fe)/N(O) abundance ratio is 
\begin{equation}
{N(Fe)\over N(O)} = {N(Fe^{i+})\over N(O^{j+})}\times 
{X(O^{j+})\over X(Fe^{i+})},
\end{equation}
where $X$ is the calculated ionic fraction in the nebula.

We first calculate the N(Fe$^{+}$)/N(O$^{0}$) ratio from the 
intensity ratio measured by OTV, $I_{Fe^{+}}(\lambda 8617)/ 
I_{O^0}(\lambda 6300)=0.069$. The particular advantage of 
ratios of the optical [Fe~II] lines and the [O~I] 6300 line 
is that the excitation energies are similar,  as are the 
critical densities $>10^7$ \cm3.
These ratios are therefore insensitive to uncertainties 
in \te \ and \ne. If one takes a density for the emitting region of 
$10^6$ \cm3 \ one gets N(Fe$^{+}$)/N(O$^{0}$)=0.069$\times$ 1.08 
=0.075. If an electron density of 4000 \cm3 were adopted instead, 
the abundance ratio would be N(Fe$^{+}$)/N(O$^{0}$)=0.069$\times$ 0.80 =
0.055, which differs by less than 30\% from the previous value. 
Similar results can be obtained using other optical [Fe~II] lines. 
A calculation of the total N(Fe)/N(O) ratio from Fe$^+$ and O$^0$ 
depends on the ionic fractions. If         
most of the emission emerges from the PIZ, where 
Fe$^+$ and O$^0$ are the dominant ionization stages, the ratio 
of the ionic fractions would be about unity and the N(Fe)/N(O) ratio (as
derived in BP95a and BPP) 
should be close to the solar value of 0.048 (Aller 1987) or 0.044 
(Seaton \etal 1994). 
If the emission originated mainly from the FIZ the ratios are
0.16, 0.27, 0.35 respectively, from
photoionization models 
from the present work (Model III), Baldwin \etal (1991), and 
Rubin \etal (1991a; 1991b). This yields lower bounds to 
the total N(Fe)/N(O) between $\sim$1/4 and 1/2 of the solar value.

Fe$^{2+}$ and O$^+$ are expected to be mostly coexisting; one can 
therefore calculate Fe$^{2+}$/O$^+$ with good accuracy. In Fig. 17 we plot 
this abundance ratio as a function of electron 
temperature and density. This was calculated 
according to Eq. (10) using line intensity ratios from OTV for 
I$_{Fe^{2+}}(\lambda 4881)$ and I$_{O^+}(\lambda 3728)$ (solid curves), and 
I$_{O^+}(\lambda 7322)$ (dashed curves). In principle, one should obtain 
the same relative abundance ratio from every line, thus the actual 
conditions of the emitting region and the abundance ratio are given by 
the point where the different curves intersect. In     
that sense, Fig. 17 shows that if the temperature of the region were 
9000 K, the electron density should be about 2500 \cm3. Similarly, 
a density of 4000 \cm3, as adopted by OTV, would yield a temperature 
close to 7000 K. We adopt mean conditions for the Fe$^{2+}$ and O$^+$ 
emitting region of \ne = 3000 \cm3 \ and \te=8000 K.
Table 7 presents the abundance ratios derived from 
individual [Fe~III] lines observed 
in Orion by OTV. The mean abundance ratio is $N(Fe^{2+})/N(O^+)= 0.011\pm      
0.003$, where the error comes from the statistical dispersion (the 
value from the blend of lines at 4986 \AA \ was excluded since it is more than 
two sigma away from the mean). The ratio of the ionic fractions, 
$X(O^{+})/X(Fe^{2+})$, according to any of our present models (III), (IV), or (V), is about 1.1,
which yields an iron to oxygen abundance ratio by number of about 
0.012$\pm$ 0.003 or 1/($3.7\pm 1.0$) of the solar value. Notice that the 
ratio of the ionic fractions from the models of Baldwin \etal (1991) and 
Rubin \etal (1991a; 1991b) are higher than the present value, 1.75 and 1.39 
respectively, and they would yield abundance ratios of about half to    
65\% solar. 

The iron to oxygen abundance ratio in the Fe~IV emitting region can be 
calculated from  the
[Fe~IV] and [O~III] lines. We use the intensity of the [Fe~IV] 2827 \AA\ 
line with respect to H$\beta$, 
as measured by Rubin \etal (1997), and OTV's observations of [O~III] 
lines also with respect to H$\beta$.
Fig. 18 shows the calculated $N(Fe^{3+})/N(O^{2+})$ abundance ratios 
as a function of temperature for several values of \ne. Here 
the solid curve represents the abundance ratio from the 
[O~III] 4363 \AA \ line and the dashed curve indicates the ratio 
obtained with the [O~III] 4959 \AA \ feature. 
As before, \te, \ne, and the abundance ratio are given
by the crossing points of the two curves.
We find the conditions for the Fe$^{3+}$ - O$^{2+}$ emitting region are 
about \te = 10500 K and \ne = 4000 \cm3, with a $N(Fe^{3+})/N(O^{2+})$ abundance 
ratio of 0.002. The uncertainty in this value from errors in the 
assumed conditions is less than 10\%, as seen from Fig. 18. 
The $X(O^{2+})/X(Fe^{3+})$ ionic fractions ratio from our present model 
(III), (IV), or (V) is 0.96, which yields a total $N(Fe)/N(O)$ 
of 0.0017 or 1/26 of the solar value. If ionic fractions from Baldwin \etal 
(1991) or Rubin \etal (1991a,b) were used, one would get a total abundance ratio of about 
1/30 of the solar value. This result is between 1.2 and 3.3 times higher than 
the values derived by Rubin \etal (1997), if one assumes a $N(O)/N(H)$ 
ratio for Orion of about 1/2 solar. Nevertheless, the present gaseous 
iron abundance from the Fe$^{3+}$ lines is about a factor of ten 
lower than our results from Fe$^+$ and Fe$^{2+}$. 
This large discrepancy exceeds the combined uncertainties of the 
observations and the atomic data and remains unexplained. 
A summary of the ionic and total Fe/O abundance ratios estimated for    
different ionization zones in Orion is presented in Table 8. 

\section{Summary and Conclusions}
        
The present study of the ionization structure and 
spectra of Fe~I-IV in Orion 
yields several conclusions that might be generally applicable to
low-ionization gaseous nebulae.
 The physical processes 
of spectral formation and photoionization
are described in the light of new atomic data for Fe~I--IV (Table 1)  that
should be sufficiently accurate for
reliable spectroscopic analysis and modeling of these ions in 
astrophysical plasmas, although more work is still needed to improve
the forbidden transition probabilities (in progress).

The study of forbidden optical emission spectra of [Fe~II],
under nebular conditions including the effects of collisional excitation, 
fluorescent excitation by UV continuum, and line 
self-absorption, reveals 
fluorescence as relatively inefficient and 
optical depth effects to be generally negligible. 
The [Fe~II] emission from H~II regions like Orion originates largely in 
high density partially ionized zones (PIZ's) within the ionization 
front. This conclusion is derived from numerous line ratio density
diagnostics, some of which are insensitive to 
fluorescent excitation. These line ratios should also be
useful for diagnostics of plasmas even when the radiation field 
is sufficiently intense to affect some of the [Fe~II] lines. 
Moreover, under fluorescent excitation 
the optical [Fe~II] emission should be accompanied by 
observable dipole allowed Fe~II lines that are absent in 
H~II regions like Orion, but are observed in 
circumstellar nebulae
and the subclass of bipolar planetary
nebulae with symbiotic star cores (a more detailed study is in
progress).

Unlike the optical emission, the [Fe~II] near-IR and IR lines can be 
easily excited at low \ne, while they are collisionally de-excited
at the high densities in the PIZ. These lines should
therefore originate from a region that extends within the lower 
density FIZ. Observational studies of the 
relative intensities of the near-IR [Fe~II] lines and their 
expansion velocities are proposed.

The [Fe~III] emission lines are primarily
collisionally excited, and the observed line ratios are 
consistent with the conditions (\te \ and \ne) of the FIZ. As expected     
from the modeled ionization structure of nebulae, there is a
correlation between the observed [Fe~III] emission and the emission 
from low ionization species like N$^+$ and O$^+$. 

The theoretical [Fe~IV] emission spectrum was studied in detail.
Under nebular conditions most of the emission 
lies in the UV, as recently observed (Rubin \etal 1997), 
while the optical emission is  
formed only by rather highly excited levels that are 
difficult to populate. 
Optical [Fe~IV] lines have been identified in only a handful of objects 
including the bipolar nebula M2-9 (Torres-Peimbert \& Arrieta 1996). 
The presence of these lines indicates unusual conditions 
for typical nebulae, with high electron densities 
up to $10^7$ \cm3  and a high degree of ionization of the plasma. 

The observed kinematics of the Orion nebula seem to be well correlated 
with the physical conditions. There is a 
distinction in the observed expansion velocities for different 
species and their degree of ionization. In particular, neutrals and 
ionized species of low ionization potential, like O$^0$, 
Fe$^+$, and Ni$^+$, which are expected to emit predominantly from the 
neutral and partially ionized zones, are distinctly separated 
by more than 10 km s$^{-1}$ from the fully ionized gas. Moreover, the 
expansion velocities of optical emission from O$^0$, 
Fe$^+$, and Ni$^+$ are remarkably similar to those of the PDR and the 
molecular core. This also provides strong evidence for the  
high densities in the PIZ. Similar kinematic analysis of other H~II 
regions is proposed.

  Photoionization modeling of Fe emission in Orion with the new atomic
data indicates that the main source of uncertainty is the assumed structure,
for example the assumption of static conditions and an {\it ad hoc}
density profile. 
Particularly problematic is the region near and within the ionization
front where neutrals and ions with low first ionization potentials emit, as illustrated
by modeling all ionization stages of oxygen
(O~I--III) in Orion simultaneously. It is found that high densities
are required at the I-front to reproduce the observed [OI] 
emission, but more realistic models would require
radiative-hydrodynamic modeling.
Also, the roles of 
enhanced electron and proton impact destruction of grains, particularly
in the high density PIZ, and the precise role of charge exchange
processes, remain to be explored.

Finally, the relative gas phase abundance of 
Fe/O in Orion is determined spectroscopically
from [Fe~II], [Fe~III], and [Fe~IV] emission separately. 
Ionic fractions of Fe and O ions that co-exist are employed 
and the physical conditions (\te, \ne) in each emitting region 
are estimated individually. This approach should be accurate
and takes into account temperature and density 
variations across the nebula. The Fe/O 
in the PIZ is found to lie between near-solar values 
down to a conservative lower limit of 1/4 solar. Taking into account 
the uncertainties, this is generally consistent with our 
previous determinations in BP95a and BPP96. The Fe/O derived from 
[Fe~III] is about 1/($4\pm 1$). For O/H in Orion of about half 
the solar value, this result agrees with most 
previous determinations of Fe/H, about one tenth solar
(e.g. OTV). In contrast, 
the Fe/O ratio obtained from [Fe~IV] 
is about 26 times lower than solar. Although considerably higher 
than the values derived by Rubin \etal \ (1997), it is still much lower 
than the determinations from other iron ions.
This apparently differential iron gas phase abundance across Orion 
obtained from 
different ions is puzzling. It seems unlikely that the 
emissivities for the [Fe~IV] UV lines could be overestimated by 
as much as a factor of 6, unless the Garstang A-values are in
substantial error (work is in progress to check these). 
However, for conditions in Orion, all    
of the emission from Fe~IV depends mostly on the collision 
strengths and only marginally on the A-values
(in the low \ne \ limit). It is noted that the 
$^4P$ and $^4D$ terms that give rise to the strongest [Fe~IV] UV lines 
are populated predominantly via collisional excitation from the 
first excited state $^4G$ which is highly metastable. Thus, one may 
expect that the emissivity of the observed lines would be very sensitive 
to radiative de-population of the $^4G$ state to the ground state, yet we 
observe that an increase of the ($^4G_J -\ ^6S_{5/2}$) A-values of {\it three 
orders of magnitude} reduces the emissivity of the 2836.6 \AA \ line 
by less than a factor of two. On the other hand, 
the new collision strengths  should be quite accurate, as discussed in 
Sections 2 and 3.5. Another possibility would be that the ionic 
fractions for the iron ions are in error. If the iron abundances 
from Fe$^{2+}$ and Fe$^{3+}$ were to be reconciled in this way the 
$X(Fe^{3+})/X(Fe^{2+})$ ionic ratio would be about 0.27 instead of the 
value of 1.8 expected from our present model, 3.4 from Baldwin \etal 
(1991), and 1.3 from Rubin \etal (1991b); the actual gas phase     
Fe/O would then be about 1/14, or Fe/H$\sim $1/28, of the solar ratios.
But such a low ratio for the 
ionic fractions would require a combined error from the photoionization 
cross sections and recombination coefficients of nearly an order of 
magnitude. This appears rather unlikely, especially since the 
unified recombination coefficients in the present work are calculated using the
same {\it
ab initio} close-coupling method as the photoionization data, ensuring
self-consistency and
minimizing the possibility of any large errors in
relative ionic fractions.
Alternatively, if the stellar radiation field were 
adjusted to reproduce this low ionic ratio, the ionic fractions for 
all other elements, e.g. C, N, O, S, etc., would also be affected.
It is rather difficult to reconcile the abundances derived from each 
ion, suggesting a gradient in the gas phase iron 
abundance in Orion via some unknown mechanism. 

\acknowledgements

We would like to thank Mike Barlow, Xiao-wei Liu, 
Don Osterbrock, and Manuel Peimbert for discussions and 
suggestions. 
We are also grateful to M\'onica Rodr\'{\i}guez for providing us with her 
observations of [FeII] and [FeIII] data.
This work was supported in part by grants from the U.S. National
Science Foundation for the Iron Project (PHY-9421898), and the NASA Long
Term Space Astrophysics program (NAS5-32643). The  computational work
was carried out on the Cray Y-MP and the massively
parallel Cray T3D at the Ohio Supercomputer Center in Columbus Ohio.


\def \aas {\it Astronomy and Astrophys. Supplement Series}
\def \rmex {Rev. Mexicana de Astronom\'{\i}a y Astrofis.}


\clearpage 

\begin{deluxetable}{lcccc}
\small
\footnotesize
\scriptsize
\tablewidth{33pc}
\tablecaption{New Atomic Data For Fe~I--V :             
(Iron Project Calculations at Ohio State)}
\tablehead{
\colhead{Ion} &       
\colhead{Photoionization} &
\colhead{Recombination} &
\colhead{Collis. Excitation} &
\colhead{Transition Prob.} }
\startdata
Fe~I & 1\tablenotemark{} & 2 & -- & 3 \nl
Fe~II & 4 & 5 & 6  & 7 \nl
Fe~III & 8 & 9 & 10 & 11 \nl
Fe~IV & 12 & 13 & 14 & 15 \nl
Fe~V & 16 & 17 & -- & 18 \nl
\enddata
\tablenotetext{}{
{\bf 1.} $\sigma_{PI}$ for 563 bound states
multiplets (Bautista \& Pradhan 1995b; Bautista 1997); 
{\bf 2.} $\alpha_R(T)$ - recombination rate
coefficients (Nahar \etal 1997); 
{\bf 3.} A-coefficients and gf-values
for 27,000 LS multiplets (Bautista 1997);
{\bf 4.} $\sigma_{PI}$ for 745 bound states (Le Dourneuf \etal 1993;
Nahar \& Pradhan 1994b);
{\bf 5.} $\alpha_R(T)$ - Nahar (1997);
{\bf 6.} $\Upsilon(T)$ - for
12,561 fine structure transitions among 142 levels (Zhang \& Pradhan 1995a);
{\bf 7.} A-coefficients and gf-values for dipole transitions among
19,267 LS multiplets and 234,689 fine structure transitions (Nahar 1995);
{\bf 8.} $\sigma_{PI}$ for 805 bound states (Nahar 1996a);
{\bf 9.} $\alpha_R(T)$ (Nahar 1996b);
{\bf 10.} $\Upsilon(T)$ for 23,871 transitions among 219 fine structure levels
(Zhang 1996);
{\bf 11.} A-coefficients and gf-values for allowed transitions among
1408 LS multiplets and 9797 fine structure transitions; A-coefficients
for 362 forbidden fine structure transitions (Nahar \& Pradhan 1996);
{\bf 12.} $\sigma_{PI}$ for 746 bound states (Bautista \& Pradhan 1997);
{\bf 13.} $\alpha_R(T)$ (Nahar, Bautista, \& Pradhan; to be submitted);
{\bf 14.} $\Upsilon(T)$ for 8,771 transitions among 140 levels
(Zhang \& Pradhan
1997);
{\bf 15.} A-coefficients and gf-values for dipole transitions among
34,635 LS multiplets (Bautista and Pradhan 1997);
{\bf 16.} $\sigma_{PI}$ for 1812 bound states (Bautista 1996);
{\bf 17.} $\alpha_R(T)$, (Nahar, Bautista \& Pradhan; to be submitted);
{\bf 18.} A-coefficients and gf-values for 129,904 LS multiplets (Bautista
1996).} 
\end{deluxetable}

\clearpage

\begin{deluxetable}{llccc}
\small
\footnotesize
\scriptsize
\tablewidth{33pc}
\tablecaption{Comparison of $\Upsilon(T=10^4$K) values for Fe~II}
\tablehead{
\colhead{Transition} &
\colhead{} &
\colhead{Present(23CC)}& 
\colhead{BP96(18CC)} &
\colhead{ZP95(38CC)} }
\startdata
$^6D_{9/2}$ & $-\ ^6D_{7/2}$ & 5.260 & 4.65 & 5.52 \nl                
            & $-\ ^6D_{5/2}$ & 1.300 & 1.29 & 1.49 \nl
            & $-\ ^6D_{3/2}$ & 0.641 & 0.813& 0.675\nl
            & $-\ ^6D_{1/2}$ & 0.297 & 0.433& 0.284\nl
            & $-\ ^4F_{9/2}$ & 1.49  & 1.31 & 3.60 \nl
            & $-\ ^4F_{7/2}$ & 0.566 & 0.614& 1.51 \nl
            & $-\ ^4F_{5/2}$ & 0.120 & 0.135& 0.497\nl
            & $-\ ^4D_{7/2}$ & 8.040 &14.30 & 10.98\nl
            & $-\ ^4D_{5/2}$ & 0.445 & 0.572& 0.560\nl
            & $-\ ^4D_{3/2}$ & 0.167 & 0.386& 0.191\nl
            & $-\ ^4P_{5/2}$ & 0.722 & 0.542& 0.948\nl
            & $-\ ^4P_{3/2}$    & 0.404 & 0.319& 0.502\nl
            & $-\ b\ ^4P_{5/2}$ & 0.273 & 0.215&0.308 \nl
            & $-\ ^4H_{13/2}$   & 0.543 & 0.196& 0.631\nl
            & $-\ ^4H_{11/2}$   & 0.256 & 0.196& 0.311\nl
            & $-\ b\ ^4H_{9/2}$ & 0.420 & 0.081& 0.402\nl
            & $-\ b\ ^4F_{7/2}$ & 0.188 & 0.039& 0.216\nl
            & $-\ ^6S_{5/2}$    & 0.857 & 0.399& ---  \nl
            & $-\ ^4G_{11/2}$   & 0.435 & 0.373& 0.344\nl
            & $-\ ^4G_{9/2}$    & 0.186 & 0.180& 0.215\nl
\enddata
\end{deluxetable}

\clearpage

\begin{deluxetable}{llccc}
\small
\footnotesize
\scriptsize
\tablewidth{33pc}
\tablecaption{Comparison of ratios of A-values for near-IR lines with observations} 
\tablehead{
 \multicolumn{2}{c}{Line ratio} & &  \\[0.2ex]
\colhead{Transitions} &
\colhead{Wavelength($\mu$m)} &
\colhead{Observed\tablenotemark{a}}&
\colhead{N \& S\tablenotemark{b}} &
\colhead{QLZ\tablenotemark{c} }}
\startdata
{$a\ ^4D_{7/2} - a\ ^6D_{9/2}\over a\ ^4D_{7/2} - a\ ^4F_{9/2}$} &
${1.257\over 1.644}$ & 1.34 & 1.36 & 1.04 \nl
{$a\ ^4D_{7/2} - a\ ^6D_{7/2}\over a\ ^4D_{7/2} - a\ ^4F_{9/2}$} &
${1.298\over 1.644}$ & 0.38 & 0.36 & 0.23 \nl
{$a\ ^4D_{7/2} - a\ ^4F_{7/2}\over a\ ^4D_{7/2} - a\ ^4F_{9/2}$} &
${1.677\over 1.644}$ & 0.13 & 0.20 & 0.41 \nl
{$a\ ^4D_{5/2} - a\ ^4F_{9/2}\over a\ ^4D_{5/2} - a\ ^6D_{5/2}$} &
${1.534\over 1.295}$ & 0.84 & 1.06 & 1.33 \nl
{$a\ ^4D_{5/2} - a\ ^6D_{3/2}\over a\ ^4D_{5/2} - a\ ^6D_{5/2}$} &
${1.328\over 1.295}$ & 0.54 & 0.61 & 0.58 \nl
{$a\ ^4D_{5/2} - a\ ^4F_{7/2}\over a\ ^4D_{5/2} - a\ ^6D_{5/2}$} &
${1.677\over 1.295}$ & 0.80 & 0.77 & 0.97 \nl
{$a\ ^4D_{3/2} - a\ ^6D_{3/2}\over a\ ^4D_{3/2} - a\ ^4F_{7/2}$} &
${1.279\over 1.599}$ & 0.72 & 0.86 & 0.73 \nl
{$a\ ^4D_{3/2} - a\ ^6D_{1/2}\over a\ ^4D_{3/2} - a\ ^4F_{7/2}$} &
${1.298\over 1.599}$ & 0.51 & 0.38 & 0.32 \nl
{$a\ ^4D_{3/2} - a\ ^4F_{5/2}\over a\ ^4D_{3/2} - a\ ^4F_{7/2}$} &
${1.712\over 1.599}$ & 0.25 & 0.26 & 0.26 \nl
{$a\ ^4D_{1/2} - a\ ^4F_{5/2}\over a\ ^4D_{1/2} - a\ ^6D_{1/2}$} &
${1.644\over 1.271}$ & 0.64 & 0.98 & 1.38 \nl
\enddata
\tablenotetext{a} {Near-IR spectra of HH1 by Everett \& Bautista (1996; 
unpublished)}
\tablenotetext{b} {Nussbaumer \& Storey (1988)}
\tablenotetext{c} {Quinet \etal (1997)} 
\end{deluxetable}

\clearpage

\begin{deluxetable}{llccc}
\small
\footnotesize
\scriptsize
\tablewidth{33pc}
\tablecaption{Comparison of ratios of A-values for optical lines with observations} 
\tablehead{
\multicolumn{2}{c}{Line ratio} & & & \\[0.2ex]
\colhead{Transitions} &
\colhead{Wavelength (\AA )} &
\colhead{Observed\tablenotemark{a}}&
\colhead{G62\tablenotemark{b}} &
\colhead{QLZ\tablenotemark{c} }}
\startdata
{$a\ ^4G_{7/2} - a\ ^4F_{3/2}\over a\ ^4G_{7/2} - a\ ^4F_{5/2}$} &
${4372.4\over 4319.6}$ & 0.83 & 0.52 & 0.51 \nl
{$a\ ^4G_{9/2} - a\ ^4F_{7/2}\over a\ ^4G_{9/2} - a\ ^4F_{5/2}$} &
${4296.9\over 4352.8}$ & 1.70 & 2.13 & 2.21 \nl
{$a\ ^4G_{9/2} - a\ ^4F_{7/2}\over a\ ^4G_{9/2} - a\ ^4F_{9/2}$} &
${4276.8\over 4177.2}$ & 2.12 & 4.53 & 3.98 \nl
{$a\ ^4G_{9/2} - a\ ^4F_{5/2}\over a\ ^4G_{9/2} - a\ ^4F_{9/2}$} &
${4352.8\over 4177.2}$ & 1.24 & 2.12 & 1.80 \nl
{$b\ ^4F_{5/2} - a\ ^4F_{7/2}\over b\ ^4F_{5/2} - a\ ^6D_{5/2}$} &
${4874.5\over 4488.8}$ & 1.17 & 1.04 & 1.35 \nl
{$b\ ^4F_{5/2} - a\ ^4F_{5/2}\over b\ ^4F_{5/2} - a\ ^6D_{5/2}$} &
${4973.4\over 4488.8}$ & 0.50 & 0.84 & 1.08 \nl
{$b\ ^4F_{5/2} - a\ ^4F_{3/2}\over b\ ^4F_{5/2} - a\ ^6D_{5/2}$} &
${5073.4\over 4488.8}$ & 0.50 & 0.39 & 0.56 \nl
{$b\ ^4F_{9/2} - a\ ^4F_{9/2}\over b\ ^4F_{9/2} - a\ ^6D_{5/2}$} &
${4814.6\over 4492.6}$ & 5.50 & 6.22 & 7.60 \nl
{$a\ ^4H_{9/2} - a\ ^4F_{7/2}\over a\ ^4H_{9/2} - a\ ^4F_{9/2}$} &
${5220.1\over 5072.4}$ & 3.40 & 4.86 & 5.15 \nl
{$a\ ^4H_{9/2} - a\ ^4F_{5/2}\over a\ ^4H_{9/2} - a\ ^4F_{9/2}$} &
${5333.7\over 5072.4}$ & 9.40 & 11.2 & 12.4 \nl
{$b\ ^4P_{5/2} - a\ ^6D_{7/2}\over b\ ^4P_{5/2} - a\ ^4F_{7/2}$} &
${4889.6\over 5432.2}$ & 3.20 & 3.69 & 2.51 \nl
{$b\ ^4P_{3/2} - a\ ^6D_{3/2}\over b\ ^4P_{3/2} - a\ ^4F_{7/2}$} &
${4774.7\over 5158.8}$ & 0.072& 0.094& 0.065\nl
{$a\ ^4H_{11/2} - a\ ^4F_{7/2}\over a\ ^4H_{11/2} - a\ ^4F_{9/2}$} &
${5261.6\over 5111.6}$ & 2.70 & 3.01 & 3.24 \nl
{$a\ ^4P_{5/2} - a\ ^4F_{7/2}\over a\ ^4P_{5/2} - a\ ^4F_{9/2}$} &
${9051.9\over 8617.0}$ & 0.21 & 0.23 & 0.24 \nl
{$a\ ^4P_{3/2} - a\ ^4F_{3/2}\over a\ ^4P_{3/2} - a\ ^4F_{7/2}$} &
${9470.9\over 8891.9}$ & 0.13 & 0.15 & 0.16 \nl
{$a\ ^4P_{1/2} - a\ ^4F_{3/2}\over a\ ^4P_{1/2} - a\ ^4F_{5/2}$} &
${9267.5\over 9033.5}$ & 1.40 & 1.26 & 1.29 \nl
\enddata
\tablenotetext{a} {Optical spectra of HH1 B\"ohm \& Solf (1990)}            
\tablenotetext{b} {Garstang (1962)} 
\tablenotetext{c} {Quinet \etal (1997)}
\end{deluxetable}

\clearpage

\begin{deluxetable}{lccccccc}
\small
\footnotesize
\scriptsize
\tablewidth{30pc}
\tablecaption{Relative intensity of [Fe~II] lines in Orion}
\tablehead{
 & \multicolumn{1}{c}{Observed} 
 & \multicolumn{3}{c}{Present\tablenotemark{e}} 
 & \multicolumn{3}{c}{B96}\\[0.2ex]
\colhead{$\lambda$ (\AA )} &
\colhead{I/I(8617)\tablenotemark{a}} &
\colhead{(I)[PIZ\%]$^f$}&
\colhead{(II)[PIZ\%]$^f$} &
\colhead{(III)}&
\colhead{(A)} &
\colhead{(B)} &
\colhead{(C)} }
\startdata
8892 & 0.19 & 0.31[32\%] & 0.33[33\%] & 0.27 &  0.31 & 0.31 & 0.36 \nl
7452 & 0.51 & 0.22[60\%] & 0.26[61\%] & 0.12 &  ---  & ---  & ---  \nl
7155 & 1.47 & 0.73[60\%] & 0.84[61\%] & 0.39 &  ---  & ---  & ---  \nl
5159\tablenotemark{b} & 1.3& 2.3[37\%]&1.4[69\%] & 1.8 &  1.7  & 1.9  & 1.1 \nl
5262 & 0.81 & 0.90[47\%] & 0.64[77\%] & 0.61 & 0.28 & 0.31 & 0.26 \nl
5334 & 0.33 & 0.43[69\%] & 0.40[87\%] & 0.18 & 0.22 & 0.24 & 0.21 \nl
4815\tablenotemark{b} & 0.94 &0.61[48\%]&0.48[72\%]& 0.41 & 1.1  & 1.2  & 0.71 \nl
4245 & 0.87 & 2.0[23\%] & 0.69[79\%] & 2.0 &  2.1  & 2.4  & 1.3  \nl
4277\tablenotemark{c} & 0.64 & 0.52[46\%] & 0.34[84\%] & 0.36 & 0.46 & 0.50 & 0.37 \nl
4287 & 1.29 & 0.31[67\%] & 0.31[79\%[ & 0.13 & ---  & ---  & ---  \nl
12567\tablenotemark{d} & 3.3-4.2&3.2[04\%] & 3.4[04\%]  & 4.1 &  3.5  & 3.4  & 2.6  \nl
 & & & & & & & \nl
\multicolumn{2}{c}{$\sigma$....}& 0.65 & 0.40 & 0.63& 0.69 & 0.80 & 0.53 \\[0.2ex]
\enddata
\tablenotetext{a}{Line intensities corrected for extinction from OTV 
except for the 12567 \AA\ line.}
\tablenotetext{b}{Unresolved [FeII] blends.}
\tablenotetext{c}{Possibly blended with OII $\lambda 4275.6$ and 4276.8 \AA }
\tablenotetext{d}{Measurements from Lowe \etal (1979; lower value) and 
Bautista \etal (1995; higher value).}
\tablenotetext{e}{(I) PIZ and FIZ with fluorescence; (II) 
PIZ and FIZ without fluorescence; (III) only FIZ with fluorescence.}
\tablenotetext{f}{The percentage contribution of the PIZ to the total intensity of 
the line is indicated brackets. The contributions of the PIZ to the 8617 \AA 
line are 24\% in model (I) and 28\% in (II).}
\end{deluxetable}

\clearpage

\begin{deluxetable}{llccccccccccc}
\small
\footnotesize
\scriptsize
\tablewidth{44pc}
\tablecaption{Optical spectrum of oxygen in Orion vs. photoionization models}
\tablehead{
 & & \multicolumn{3}{c}{Observed/I(H$\beta)\times 100$}& \multicolumn{3}{c} 
{B96} & \multicolumn{5}{c}{Present\tablenotemark{b}} \\
[0.2ex]
\colhead{Ion} &
\colhead{line (\AA ) } &  
\colhead{OTV}&
\colhead{B96 } & 
\colhead{G94\tablenotemark{a} } &
\colhead{(A)} &
\colhead{(B)} &
\colhead{(C)} &
\colhead{(I)} &
\colhead{(II)} &
\colhead{(III)} &
\colhead{(IV)} &
\colhead{(V)} }
\startdata
[O~I] & 6300 & 0.959 & 0.722 & 0.70 & 0.341 & 0.336 & 0.699 & 0.34 & 0.20 & 
0.17 & 0.75 & 0.91 \nl
[O~I] & 5577 & 0.058 &$<$ 0.0136& ---  & 0.0044& 0.0043& 0.0095& 0.004& 0.002&
0.002& 0.006& 0.009\nl
[O~II]& 3727 & 146   & 94    & 123  & 188   & 188   & 149   & 186  & 209  &
145  & 126  & 132  \nl
     & 7320 & 6.21 & ---   & 7.10  & ---  & ---   & ---   & 13.8  & 7.86 & 
4.44 & 11.6 & 9.19 \nl
     & 7330 & 5.47 & ---   & 5.92  & ---  & ---  & ---  & 11.2 & 6.35 &
3.56 & 9.40 & 7.45 \nl
[O~III]&4363 & 1.39 & ---  & 1.00 & ---  & ---  & ---  & 1.93 & 2.39 & 
0.74 & 0.69 & 0.79 \nl
     & 4959 & 100.2& ---  & 104.8& ---  & ---  & ---  & 132  & 169  &
87.8 & 75.9 & 87.6 \nl
     & 5007 & 302  & 343  & ---  & 465  & 460  & 379  & 395  & 507  &
263  & 228 & 263  \nl
\enddata
\tablenotetext{a}{Greve \etal (1994)}
\tablenotetext{b}{Results from present photoionization models including 
the contribution from a high density ionization front (see text for 
an explanation of the different models)}
\end{deluxetable}

\clearpage

\begin{deluxetable}{cccc}
\small
\footnotesize
\scriptsize
\tablewidth{34pc}
\tablecaption{Fe$^{2+}$/O$^+$ abundance ratios in Orion}
\tablehead{
\colhead{$\lambda $(\AA )} &     
\colhead{Transition} & 
\colhead{$I_{Fe^{2+}}(\lambda )/I_{0^+}(3727)$} &
\colhead{$N(Fe^{2+})/N(O^+)$ }} 
\startdata
5412 & $a\ ^5D_1-a\ ^3P_2$ & $2.4\times 10^{-4}$ & 0.0084 \nl
5270 & $a\ ^5D_3-a\ ^3P_2$ & $3.7\times 10^{-3}$ & 0.012 \nl
4987 & $a\ ^5D_4-a\ ^3H_6$ & $2.3\times 10^{-4}$ & 0.003 \nl
     &+$a\ ^5D_3-a\ ^3H_4$ &                     &       \nl
4881 & $a\ ^5D_4-a\ ^3H_4$ & $2.9\times 10^{-3}$ & 0.0092\nl
4658 & $a\ ^5D_4-a\ ^3F_4$ & $9.2\times 10^{-3}$ & 0.036 \nl
4702 & $a\ ^5D_3-a\ ^3F_3$ & $1.7\times 10^{-3}$ & 0.0099\nl
4734 & $a\ ^5D_2-a\ ^3F_2$ & $5.6\times 10^{-4}$ & 0.0095\nl
4607 & $a\ ^5D_4-a\ ^3F_3$ & $4.2\times 10^{-4}$ & 0.013 \nl
4755 & $a\ ^5D_3-a\ ^3F_4$ & $1.2\times 10^{-3}$ & 0.012 \nl 
4769 & $a\ ^5D_2-a\ ^3F_3$ & $6.3\times 10^{-4}$ & 0.011 \nl
4778 & $a\ ^5D_1-a\ ^3F_2$ & $4.5\times 10^{-4}$ & 0.016 \nl
\enddata
\end{deluxetable} 

\clearpage
 
\begin{deluxetable}{cccc}
\small
\footnotesize
\scriptsize
\tablewidth{36pc}
\tablecaption{Fe/O abundance ratios in Orion}
\tablehead{
\colhead{Zone} &
\colhead{Fe$^{i+}$/O$^{j+}$ } &
\colhead{Fe/O} &
\colhead{ (Fe/O)/(Fe/O)\sun  } }
\startdata
$Fe^+-O^0$    & 0.065 & 0.010 - 0.065 & 1/4 - 3/2 \nl
$Fe^{2+}-O^+$ & 0.014$\pm $ 0.004 & 0.015$\pm $ 0.005 & $1/(3.7\pm 1.0)$ \nl
$Fe^{3+}-O^{2+}$& 0.0020$\pm $ 0.0006& 0.0017$\pm $ 0.0005&$1/(26\pm 8)$\nl
\enddata
\end{deluxetable}

\clearpage

\section*{Figure Captions}

\noindent{\bf Figure 1.}
Photoionization cross sections for the ground states
of Fe~I--V including detailed autoionization resonances. Also plotted are the
results of Reilman \& Manson (1979; square dots) and Kelly (1972; dashed 
line).

\noindent{\bf Figure 2.}
New recombination rate coefficients for Fe~I (Nahar, Bautista,
\& Pradhan 1997), Fe~II (Nahar 1997), and Fe~III (Nahar 1996b). These
rates are compared with previous dielectronic plus radiative recombination 
results by Woods \etal (1981; dashed line).

\noindent{\bf Figure 3.}
Energy diagram of Fe~II with infrared and optical lines considered.

\noindent{\bf Figure 4.}
[Fe~II] line ratios vs. log \ne (\cm3) for \te = 10000 K. The different 
curves represent pure collisional excitation (solid), collisional 
and fluorescent excitation without optical depth effects (dotted), 
collisional 
and fluorescent excitation including line self-shielding (long dashed), 
and collisional 
and fluorescent excitation for a UV field ten times that in Orion 
(short dashed). Collisionally excited line ratios
calculated with collision strengths of the present 
23CC calculation (short-dash and dot curves) are also shown.  
The predicted line ratios by Baldwin \etal (1996) 
are indicated by square dots. The horizontal lines indicate the 
observed values by OTV (solid)
and Rodr\'{\i}guez (1996; dashed lines).

\noindent{\bf Figure 5.} 
Similar to Fig. 4 but, with optical measurements by OTV and 
near-IR observations by Lowe \etal (1979; dotted line) and 
Bautista \etal (1995; solid line).

\noindent{\bf Figure 6.}
Energy diagram of Fe~III with infrared and optical lines considered.

\noindent{\bf Figure 7.}
[Fe~III] line ratios vs. log \ne (\cm3) for \te = 9000 K. 
The horizontal lines indicate the 
observed values in Orion by OTV (solid) and the range of values by Greve \etal (1994; dotted lines), 
and Rodr\'{\i}guez (1996; dashed lines) for optical lines 
and Bautista \etal (1995; dotted lines) and DePoy \& 
Pogge (1994; dashed lines) for the near-IR lines.

\noindent{\bf Figure 8.}
Energy diagram of Fe~IV with optical and ultraviolet lines considered.

\noindent{\bf Figure 9.}
[Fe~IV] line ratios vs. log \ne (\cm3) for \te = 9000 K.

\noindent{\bf Figure 10.}
Line ratio \ne \ diagnostics from [Fe~IV] and [O~III] optical lines 
of high density plasma in the planetary nebula with a symbiotic star 
core M2-9. Observations by Torres-Peimbert \& Arrieta (1996).

\noindent{\bf Figure 11.}
[O~I] $\lambda\lambda 6300+6364$ to $\lambda5578$ line ratio
vs. log N$_e$(cm$^{-3}$) for T = 5000, 8000, 9000, 10000, and 20000 K. The value
of this line ratio reported by OTV and the upper limits given
by Baldwin \etal (1996; HST and CTIO) are represented by horizontal
dashed lines.

\noindent{\bf Figure 12.}
The observed velocities of optical lines in Orion vs. the minimum 
photon energy required to produce the ionized specie (adapted from 
Kaler 1967 and Balick \etal 1974). The velocities of the molecular 
cloud (OMC-1) and the photodissociation region (PDR) are also indicated. 
The observations are from Kaler (1967; empty squares), Fehrenbach 
(1977; filled circles), and O'Dell \& Wen (1992; filled squares).

\noindent{\bf Figure 13.} 
Computed physical conditions of a constant gas pressure cloud as a function of the 
distance from the illuminated face.

\noindent{\bf Figure 14.}
Sample of ionizing fluxes vs. the photon energy near the illuminated 
face (a) and near  one half thickness of the cloud (b). 
The ionization limits for Fe~I--III are also indicated.

\noindent{\bf Figure 15.}
Computed photoionization rates in arbitrary units for Fe~I--III as a function of the 
distance from the illuminated face of the cloud. The solid curves 
represent the results using the new photoionization cross sections. These 
are compared with the results obtained with cross sections by 
Reilman \& Manson (1979; dotted curves) and Kelly (1972; dashed curve).

\noindent{\bf Figure 16.}
Computed ionization structure of iron in a constant gas pressure cloud as a function of 
the       
distance from the illuminated face. The results with the new photoionization-recombination 
data (solid curves) are compared with the results from CLOUDY (dotted 
curves). 

\noindent{\bf Figure 17.}
The Fe$^{2+}$/O$^+$ abundance ratio in Orion as a function of the 
assumed temperature and electron density of the region. The 
line intensities are from OTV for
I$_{Fe^{2+}}(\lambda 4881)$ and I$_{O^+}(\lambda 3728)$ (solid curves) and
I$_{O^+}(\lambda 7322)$ (dashed curves).

\noindent{\bf Figure 18.}
The Fe$^{3+}$/O$^{2+}$ abundance ratio in Orion as a function of the 
assumed temperature and electron density of the region. The 
line intensities are from Rubin \etal (1997) for the 
[Fe~IV] 2827 \AA line and from OTV for the [O~III] 4363 \AA \ (solid line) 
and the [O~III] 4959 \AA \ line (dashed line).

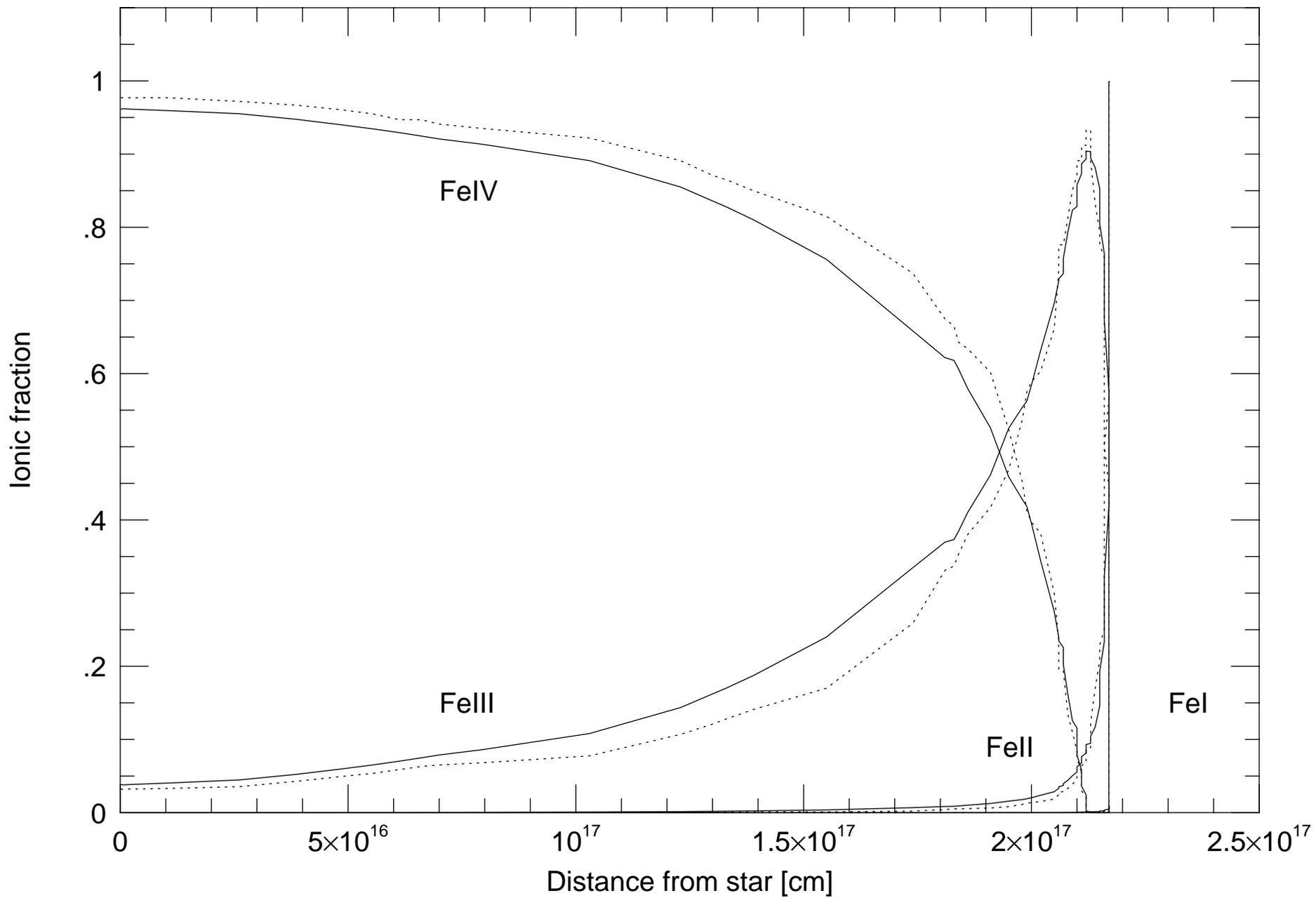
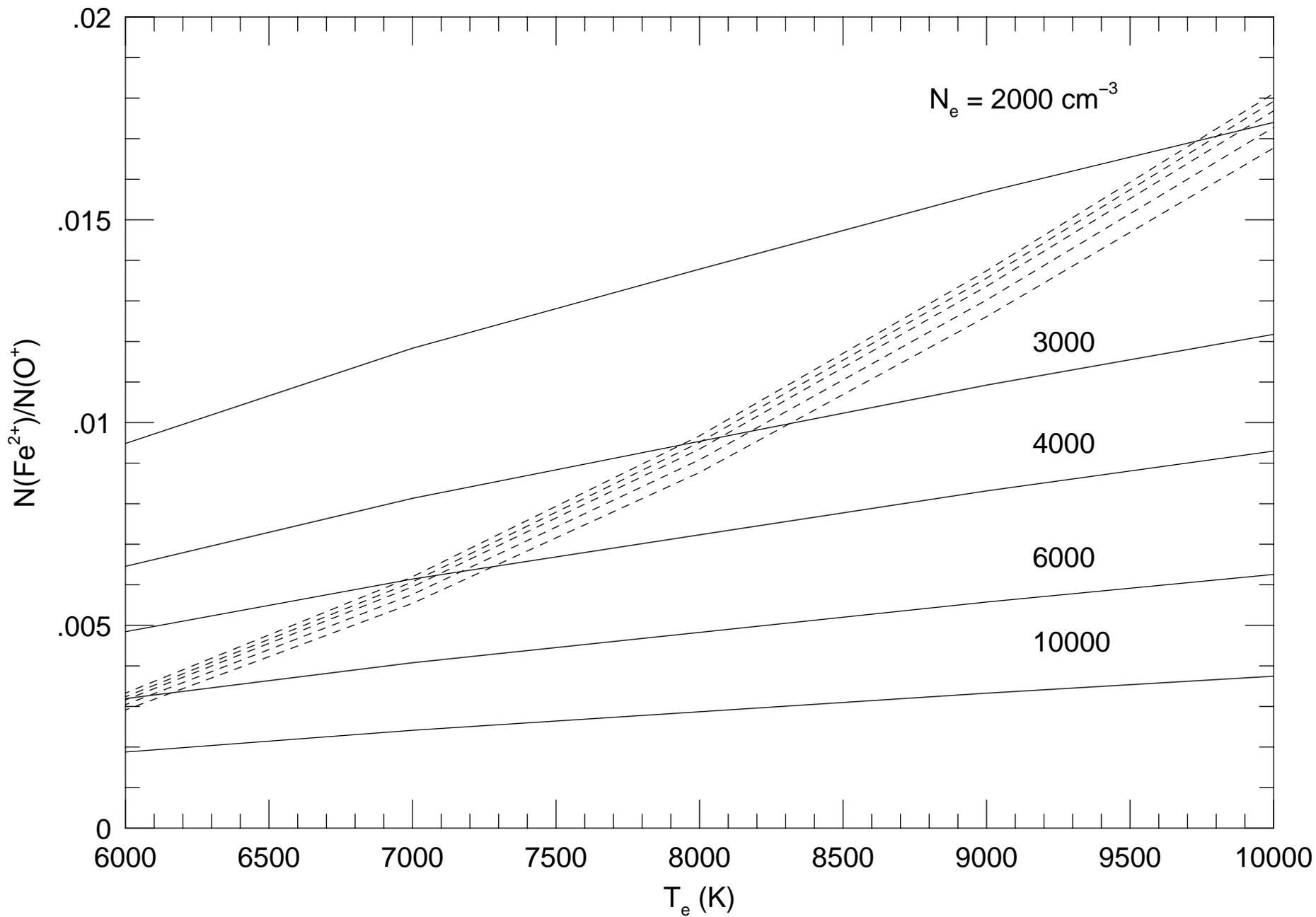
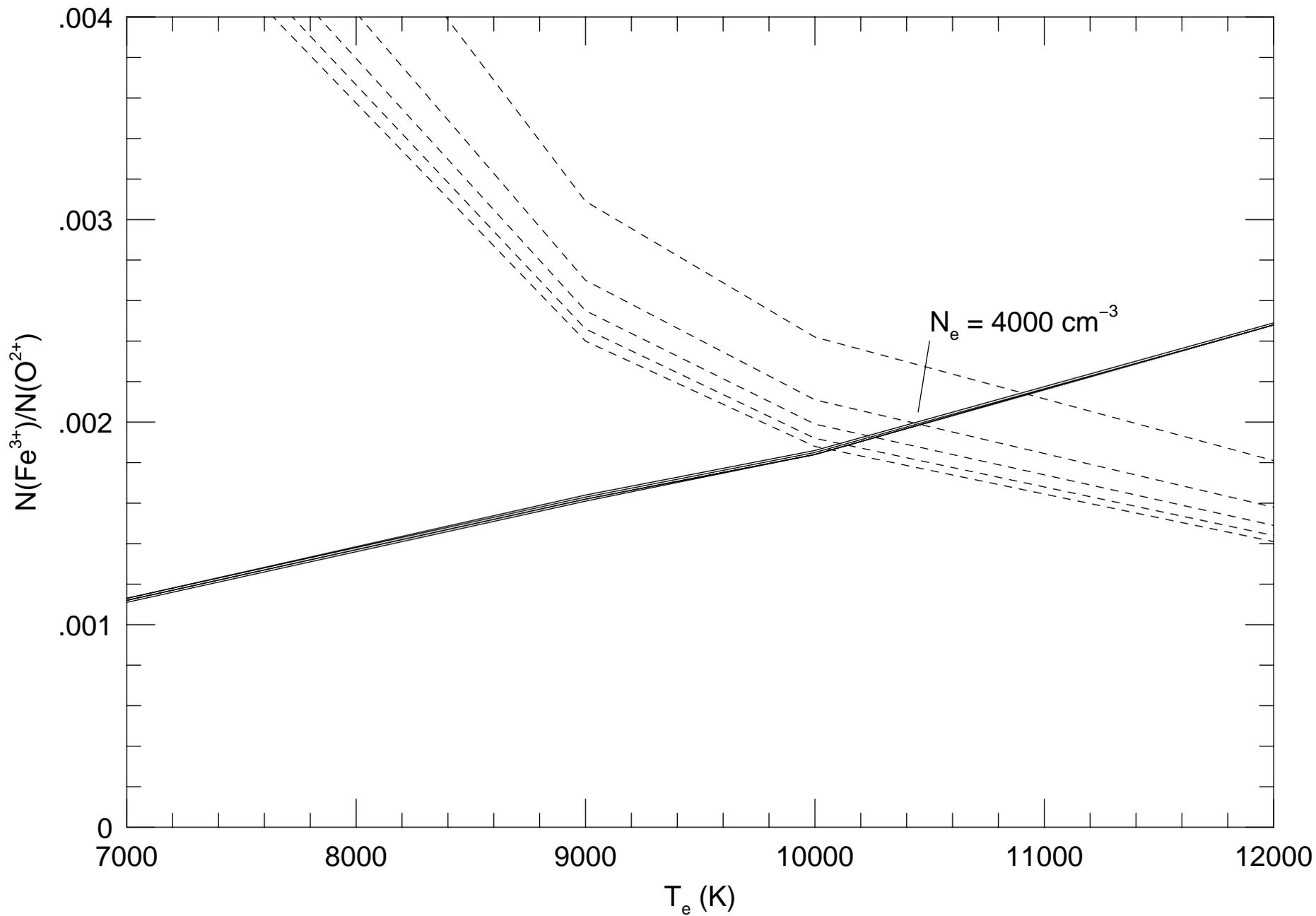
\end{document}